\documentclass[usenatbib]{mn2e}
\usepackage{epsfig}
\usepackage{amsmath}
\usepackage{ulem}
\usepackage{times}

\def\gsim{\mathrel{\raise0.35ex\hbox{$\scriptstyle >$}\kern-0.6em 
\lower0.40ex\hbox{{$\scriptstyle \sim$}}}}
\def\lsim{\mathrel{\raise0.35ex\hbox{$\scriptstyle <$}\kern-0.6em 
\lower0.40ex\hbox{{$\scriptstyle \sim$}}}}

\date{\today}
\title[A H$\alpha$\ and MIR imaging of a $z= 0.8$ cluster]
{Panoramic H{\boldmath $\alpha$} and mid-infrared mapping of star formation 
in a {\boldmath $z=0.8$} cluster} 

\author[Y. Koyama et al.]{
\parbox[t]{\textwidth}{
Yusei Koyama$^{1}$\thanks{E-mail: koyama@astron.s.u-tokyo.ac.jp},
Tadayuki Kodama$^{2}$,
Kazuhiro Shimasaku$^{1,3}$,
Masao Hayashi$^{1}$, 
Sadanori Okamura$^{1,3}$, 
Ichi Tanaka$^{4}$ and
Chihiro Tokoku$^{5}$ 
}
\vspace*{6pt}\\
$^{1}$Department of Astronomy, School of Science, The University of Tokyo,
Tokyo 113-0033, Japan\\
$^{2}$National Astronomical Observatory of Japan, Mitaka, Tokyo
181-8588, Japan\\
$^{3}$Research Center for the Early Universe, School of Science, The
University of Tokyo, Tokyo 113-0033, Japan \\
$^{4}$Subaru Telescope, National Astronomical Observatory of Japan, 650 
North A'ohoku Place, Hilo, HI 96720, USA\\
$^{5}$Astronomical Institute, Tohoku University, Aramaki, Aoba-ku, Sendai,
Miyagi 980-8578, Japan
}
\begin{document}

\maketitle

\begin{abstract}
We present the first wide-field H$\alpha$\ imaging survey
around the distant cluster RXJ1716.4+6708 at $z=0.81$ with a narrow-band
filter on MOIRCS/Subaru, which reveals the star formation activities
down to a star formation rate (SFR) of $\sim$1$M_{\odot}$/yr without 
extinction correction. Combining with a wide-field mid-infrared (MIR) 
imaging survey with AKARI satellite, we compare in detail the unobscured
and obscured star formation activities in the cluster. We find 
that both H$\alpha$\ emitters and MIR galaxies avoid the cluster central 
region and their spatial distribution is quite similar. 
Most of the H$\alpha$\ emitters show blue colours, but we find 
some H$\alpha$\ emitters on the red sequence. The MIR galaxies 
tend to be systematically redder than the H$\alpha$\ emitters probably 
due to heavy dust extinction. Interestingly, the red H$\alpha$\ 
emitters and the red MIR galaxies (i.e. dusty red galaxies) are most 
commonly seen in the medium-density environment such as 
cluster outskirts, groups and filaments, where optical colours of 
galaxies change. 
We investigate the amount of hidden star formation by calculating
a ratio, SFR(IR)/SFR(H$\alpha$), and find that $A_{\textrm{H}\alpha}$ 
exceeds $\sim$ 3 in extreme cases for actively star-forming galaxies 
with SFR(IR)$\gsim$20$M_{\odot}$/yr. It is notable that most of 
such very dusty galaxies with $A_{\textrm{H}\alpha}$$\gsim$3 are 
also located in the medium-density environment. 
These findings suggest that dusty star formation
is triggered in 
the in-fall region of the cluster, implying a probable link 
between galaxy transition and dusty star formation. 
We finally calculate the cluster total SFR and find that 
the cluster total SFR based on H$\alpha$\ alone can be 
underestimated more than factor $\sim$2 even after 1 mag 
extinction correction.  We suggest that the 
mass-normalized cluster SFR rapidly declines since $z\sim 1$ 
following $\propto (1+z)^6$, although the uncertainty is still large.   
\end{abstract}
\section{Introduction}
\label{sec:intro}
It is well known that properties of galaxies are strongly correlated
with environment in the sense that galaxies in high-density environment
tend to be red and have early-type morphology. 
This is first quantitatively noted by \cite{dre80} in nearby clusters,
and many other studies have confirmed such correlations as well as
extended them in redshift space (e.g. \citealt{pos84}; \citealt{dre97};
\citealt{got03}; \citealt{pos05}). 
Recent large, intensive spectroscopic surveys for the local Universe
show that star-forming activity is also a strong function of environment
(e.g. \citealt{lew02}; \citealt{gom03}; \citealt{tan04}) in the sense
that star-forming activity is weaker in the higher-density environment.
Cluster galaxies are in fact mostly red early-types with little on-going 
star formation at least in the local Universe.
However, if we look at distant clusters, we can find many blue 
star-forming galaxies in cluster environment (the so-called 
Butcher-Oemler effect: \citealt{but84}).
Therefore by observing distant clusters at various redshifts 
(i.e. clusters in the past Universe at various epochs)
we should be able to identify when and how the activities of cluster
galaxies are changed from blue star-forming populations to red quiescent ones.

The 'hidden' star forming galaxies in clusters are also key population
to characterize the evolution of cluster galaxies.
Recent mid-infrared (MIR) observations of distant clusters 
with space telescopes such as 
Spitzer and AKARI discovered many dusty galaxies in $z\gsim 0.5$ clusters 
(\citealt{gea06}; \citealt{mar07}; \citealt{bai07}; \citealt{sai08}; 
\citealt{koy08}; \citealt{kri09}).  Although the number of clusters 
studied in MIR is still small, \cite{sai08} showed the trend 
that the dusty star-forming population is more commonly seen in the more 
distant clusters (see also \citealt{hai09b}).  \cite{koy08} showed 
the possibility of environmental dependence in the fraction of 
dusty star-forming galaxies at $z\sim 0.8$, and suggested
that the MIR activity is enhanced in the 'medium-density' environment such 
as cluster outskirts, groups and filaments, where galaxy colours
sharply change from blue to red.

Another good tracer of star formation is optical emission lines
from galaxies.  In particular, H$\alpha$\ line (6563\AA) is considered
to be one of the best star formation indicators, since it directly reflects 
the emission from H{\sc ii} region in the star-forming sites.
Moreover, H$\alpha$ line is less affected by dust extinction or metallicity, 
compared to [OII] line which has been commonly used in the studies of distant 
galaxies.
However, since H$\alpha$\ line shifts to near-infrared (NIR)
regime at $z\gsim 0.4$, it has been difficult to conduct a large H$\alpha$\
study for $z\gsim 0.4$ clusters until recently.  There are some
H$\alpha$\ studies for low-$z$ clusters with optical spectroscopy 
(\citealt{cou01}; \citealt{bal02}) and with narrow-band 
imaging (\citealt{bal00}; \citealt{kod04}), but the number of known 
clusters studied in H$\alpha$\ is really small.  For distant clusters,
\cite{fin04} and \cite{fin05} conducted narrow-band H$\alpha$\ 
imaging surveys for several $z\sim$0.7--0.8 clusters in NIR. 
However, their field coverage is limited to the central regions of
the clusters.  As suggested in \cite{koy08}, cluster surrounding environment
is likely to be the key environment for galaxy evolution, but
such spatially extended regions have been much less explored mainly
due to the limited field of view of NIR instruments.

It is thus ideal to observe distant clusters in both of the two major,
but independent indicators of star formation, namely, H$\alpha$\ and MIR,
over a wide area from cluster cores to surrounding environments.
It is also important to quantify any different views of star forming
activities seen by the two different indicators.
Mapping out dusty and non-dusty star formation activities in
distant clusters may give us a clue to understand what is actually
happening in the transition environments.
In this paper, we present the first such wide-field H$\alpha$\ imaging
survey over the known structures including the central cluster
RXJ1716.4+6708 at $z\sim 0.8$.
Combining with the similarly wide-field MIR data of this cluster
taken with AKARI, we also attempt to compare the H$\alpha$\ and MIR 
views of the $z >0.8$ cluster.

The RXJ1716.4+6708 cluster (hereafter RXJ1716) that we target in this
paper is a rich and probably unvirialized cluster at $z=0.81$.
The cluster was first discovered in the ROSAT North Ecliptic Pole Survey 
(NEP: \citealt{hen97}).  Optical spectroscopy was performed by 
\cite{gio99} and 37 cluster members were identified.
Using this spectroscopic sample, \cite{gio99} determined the 
cluster redshift $z_{cl}=0.809$ and the velocity dispersion  $\sigma = 
1522 ^{+215} _{-150}$~km s$^{-1}$.  This velocity dispersion is 
relatively large for its rest-frame X-ray luminosity of
$L_{bol} = 13.86 \pm 1.04  \times 10^{44} $erg s$^{-1}$ and 
the temperature $kT = 6.8 ^{+1.0}_{-0.6}$~keV (\citealt{ett04}). 
This indicates that this cluster is not virialized yet and in the process of
active assembly.  In fact, this cluster has a small subcluster or group to 
the northeast of the main cluster.  The morphology of this cluster 
in a X-ray image 
elongates towards the direction of the subcluster (e.g. \citealt{jel05}).  
\cite{koy07} performed a wide-field, multi-colour optical imaging of this
cluster and discovered prominent large-scale structures 
penetrating the cluster core and the second group of this 
cluster towards the southwest of the cluster core, based on 
the photometric redshift technique.  
The weak-lensing mass of this cluster is estimated to be 
$2.6 \pm 0.9 \times 10^{14} h^{-1} M_{\odot}$ (\citealt{clo98}). 
This is consistent with the estimated mass based on the
X-ray data in \cite{ett04};
$M_{\rm{tot}} = 4.35 \pm 0.83  \times 10 ^{14} M_{\odot}$. 

The structure of this paper is as follows.  In Section~2, we summarize 
our optical, NIR and MIR data. In Section~3, we show the selection
technique of H$\alpha$\ emitters from cluster members. The derivation 
of H$\alpha$-derived star-formation rates is shown in Section~4.
We show the results and discussions from Section~5 to Section~8, 
and summarize our results in Section~9. 
Throughout this paper, we use $\Omega_M =0.3$, $\Omega_{\Lambda} =0.7$, 
and $H_0 =70$ km s$^{-1}$Mpc$^{-1}$.  Magnitudes are all given in the 
AB system.

\section{Data}
\label{sec:data}
\subsection{Optical data}
\label{subsec:optical_data}

We use the optical data in \cite{koy07}.
We performed deep and wide-field $VRi'z'$ imaging of the RXJ1716
cluster with Suprime-Cam (\citealt{miy02}) on the Subaru Telescope
(\citealt{iye04}), as a part of the PISCES project
(Panoramic Imaging and Spectroscopy of Cluster Evolution with 
Subaru; \citealt{kod05}). 
A summary of the data and the method of data reduction is described 
in \cite{koy07}, and we just repeat some important points here.

We reduced the data using a pipeline software SDFRED (\citealt{yag02}; 
\citealt{ouc04}).  Source detection and photometry were done 
using SExtractor software (\citealt{ber96}).  We use 2$''$ aperture 
magnitudes (MAG\_APER) for measuring galaxy colours and MAG\_AUTO are used as 
total magnitudes of galaxies.  Our optical catalogue was constructed 
for objects brighter than $z'= 24.9$ which corresponds to the 5$\sigma$
detection limit in $z'$-band.
For our use in this paper, we applied a small aperture correction 
to the aperture magnitudes.  This is because the seeing size of our 
near-infrared (NIR) data is slightly larger (0.89$''$, see
Section~2.2.1) than that of the optical images ($\sim 0.7''$).
We smoothed our optical images to $0.89''$ and determined the correction
value, but the correction is small ($\sim$0.02 mag).
We estimated photometric redshifts (phot-$z$) for all the galaxies
using the phot-$z$ code of \cite{kod99}.  
We found that $> 90$\% of spectroscopically confirmed members 
listed in \cite{gio99} fall within the redshift range of
$0.76\le z_{\textrm{phot}} \le 0.83$ (see fig.2 of \citealt{koy07}),
and used this range to select member candidates.

\subsection{Near-infrared(NIR) data}
\label{subsec:NIR_data}

\subsubsection{Observation and data reduction}

We observed the RXJ1716 cluster in $J$ and NB119 filter
with MOIRCS on the Subaru Telescope (\citealt{ich06}; \citealt{suz08}). 
The NB119 filter perfectly matches to H$\alpha$\ lines at $z=0.81$
(i.e., the redshift of our target cluster).
In Fig.~1 we compare the velocity distribution of spectroscopically
confirmed members listed in \cite{gio99} with the NB119 filter response
function. 
It clearly shows that we can detect more than $\sim$80\% of H$\alpha$
emitting members within the FWHM of the NB119 filter if they have
the same velocity distribution as that of the spectroscopic members
in \cite{gio99} and if their H$\alpha$\ EWs are larger than the threshold
of our observation. 
The $J$-band filter samples mainly continuum flux.
Therefore, with a combination of $J$ and NB119 filters, we can conduct
an unbiased, deep H$\alpha$\ imaging survey of this cluster. 
To neatly cover the large-scale structures discovered by \cite{koy07}, 
we set eight fields of view of MOIRCS as shown in Fig.~2.  It can be seen 
that F1, F2, F7 and F8 have the full instrument coverage of
4$' \times$7$'$ (i.e. the original FoV of MOIRCS), and all the other
fields, F3, F4, F5 and F6, have only a half size. Unfortunately,
at the time of observation, one of the two MOIRCS chips was replaced
with an engineering-grade one due to the instrumental failure.
We should still stress, however, that this is the first H$\alpha$\ 
imaging survey covering such a wide range in environment at $z\gsim 0.8$. 

The data is reduced in a standard manner primarily using a pipeline
software MCSRED (Tanaka et al. in prep).
Some frames taken with NB119 show a strong fringe pattern, 
and we used a self-made fringe subtraction software.  
After matching the source positions on individual frames, we coadded to make
a final image for each field, and then mosaiced all the images together.
A summary of the $J$ and the NB119 data is shown in Table~1. 
We match the seeing size of all the images to 0.89$''$,
which is the worst seeing size in the $J$-band.
The photometric zero-points are derived using some standard stars observed 
on 27th May 2007. 
We properly scale F3--F8 data to F1 or F2 using the photometry of 
the same objects in the overlapping regions, although the amount of such
scaling is small ($<$0.1 mag).  
We finally check that there is no significant difference in the zero-points
between our FoVs using stars in the 2MASS catalogue
(\citealt{jar00}) that fall within our observed fields.
The Galactic extinction is corrected using the dust map by \cite{sch98}.
Limiting magnitudes are estimated by measuring 
the scatter in fluxes in randomly distributed 2$''$ apertures 
for each FoV. Note that some overlapped regions are deeper.
 
 \begin{figure}
   \begin{center}
    \leavevmode
    \vspace{-1.2cm}
    \rotatebox{0}{\includegraphics[width=8.5cm,height=8.5cm]{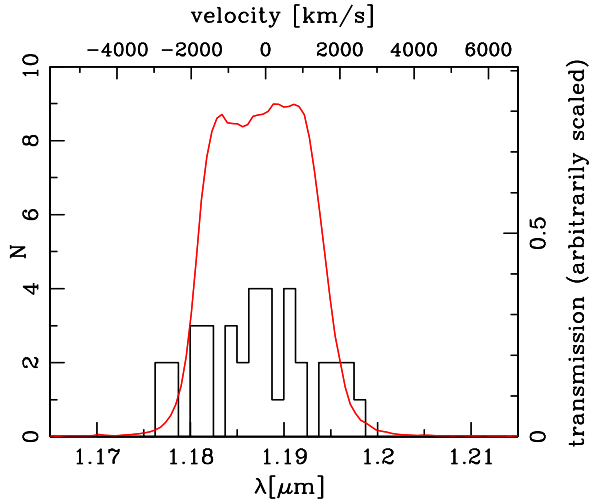}}
    \vspace{-0.7cm}
   \end{center} 
   \caption{The transmission curve of the NB119 filter. The histogram 
 shows the radial velocity distribution of spectroscopically confirmed members
 listed in Gioia et al. (1999), with respect to the velocity centre of the cluster.
 The wavelength of the histogram
 indicates the position where H$\alpha$\ line of each member galaxy falls.}
\label{fig:NB119_trans}
 \end{figure}
 \begin{figure}
   \begin{center}
    \leavevmode
    \rotatebox{0}{\includegraphics[width=8.5cm,height=8.5cm]{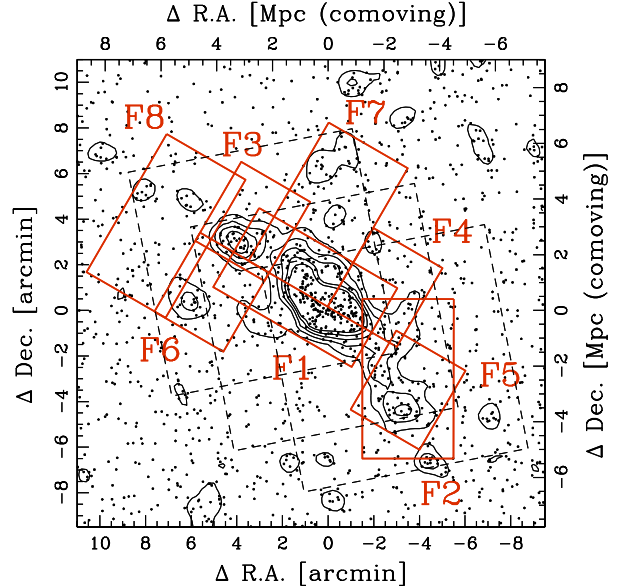}}
    \vspace{-0.7cm}
   \end{center} 
   \caption{Our MOIRCS FoVs (F1--F8) are shown by the solid-line boxes. 
  The three large dashed-line squares indicate the FoVs of our MIR
  observations with AKARI satellite (10$' \times$ 10$'$ for each). 
  The phot-$z$ selected member galaxy candidates in Koyama et al. (2007)
  are shown by the small dots.
  The contours indicate 1,2,3,4,5$\sigma$ above the average local number 
  density of the phot-$z$ selected member candidates. }
\label{fig:FoVs}
 \end{figure}
\begin{table*}
\begin{tabular}{p{8mm}|p{30mm}|p{15mm}|p{15mm}|p{12mm}|p{12mm}|p{12mm}|p{12mm}|p{12mm}|p{12mm}}
\hline
\hline
field  & Obs. date &  FoV  &   chip   &  \multicolumn{2}{c}{Exp. time
 (sec)}  & \multicolumn{2}{c}{Limit mag. (2$''$, 5$\sigma$)} &  
 \multicolumn{2}{c}{PSF FWHM}  \\
       &         & (arcmin$^2$) &       &  $J$  & NB119  &  $J$  & NB119 &
 $J$  &  NB119  \\  
\hline
F1  &  2007/05/27, 2009/09/07 &  4$' \times$ 7$'$   &  1 and 2 &
3000  &  6000  &  23.5  &  23.3  &  0.6$''$--0.8$''$ & 0.8$''$  \\
F2  &  2007/05/27, 2009/09/07 &  4$' \times$ 7$'$  &  1 and 2 &
2040  &  5100  &  23.5  & 23.3 &  0.6$''$ & 0.8$''$  \\
F3  &  2008/04/19, 2008/06/30  &  4$' \times$ 3.5$'$  &  2  &
4170 & 4800  &  23.6 &  23.6 & 0.5$''$--0.8$''$ & 0.5$''$  \\
F4  &  2008/06/29  &  4$' \times$ 3.5$'$    &  2  &
3840 & 6000 &  23.6 &  23.4  &  0.7$''$ &  0.6$''$ \\
F5  &  2008/04/21, 2008/06/29 & 4$' \times$ 3.5$'$   & 2 & 
3960 & 4500 & 23.4 & 23.3 & 0.6$''$--0.9$''$ & 0.6$''$ \\
F6  &  2008/04/20, 2008/06/30 &  4$' \times$ 3.5$'$  & 2 & 
3210 & 6000 & 23.4 & 23.5 & 0.6$''$--0.8$''$ & 0.6$''$  \\
F7  &  2009/09/03, 2009/09/04  &  4$' \times$ 7$'$   &  1 and 2 &
2420  &  4800  &  23.6  &  23.2  &   0.6$''$--0.9$''$  & 0.8$''$  \\
F8  &  2009/09/04, 2009/09/05 &  4$' \times$ 7$'$  &  1 and 2 &
2340  &  4800  &  23.6  & 23.2 &  0.5$''$ & 0.6$''$  \\

\hline 
\end{tabular}
\caption{A summary of our near-infrared imaging data in $J$ and NB119.  }
\label{tab:data_summary}
\end{table*}

\subsubsection{Photometric catalogue}
We first construct an object catalogue in $J$ and NB119 with the 
SExtractor (\citealt{ber96}) using its two-image mode with the 
NB119 image for source detection.
We use MAG\_APER (2$''$ aperture magnitude) for
colours of galaxies and MAG\_AUTO for total magnitudes. 
As shown in Table.~1, the depths of the data vary from field to field.
We include objects with NB119 $< 23.2$, which is $> 5\sigma$ detection 
in all the fields. We also require $J< 23.9$ mag ($\sim $3$\sigma$ in
$J$-band) for our sample to avoid spurious detection in NB119. 
After matching to the optical catalogue constructed by \cite{koy07}, we 
construct a $J$- and NB119-detected catalogue of the RXJ1716 cluster. 
Stars and objects around very bright stars are rejected through this process. 
The catalogue contains a full range of information of the NB119-detected
galaxies, including photometries in $VRi'z'J$ and NB119 bands as well as
the phot-$z$ values from \cite{koy07}.

\subsection{Mid-infrared(MIR) data}
\cite{koy08} observed this cluster at 15$\mu$m (L15 filter) with 
infrared camera (IRC)
onboard the AKARI satellite (\citealt{ona07}; \citealt{mur07}).
The FoVs of our MIR observations are shown with dashed-line squares
in Fig.~\ref{fig:FoVs}.  It is notable that the areas of our NIR observations
are mostly covered by those of MIR observations. 
Our 15$\mu$m filter captures the peaks of Polycyclic Aromatic
Hydrocarbon (PAH: \citealt{pug89}) emissions at 
rest-frame 7.7 and 8.6~$\mu$m at $z\sim 0.8$, which are considered to be
a good indicator of dusty star formation (e.g. \citealt{cha01}). 
The observed 15~$\mu$m flux (FLUX\_AUTO value from SExtractor) 
is first converted to the total IR luminosity, $L$(IR), using a 
relation between $\nu L_{\nu, 8\mu \textrm{m}}$ and $L$(IR) derived 
from the starburst SEDs in \cite{lag04} (see fig.4 of \citealt{koy08}). 
Thus derived $L$(IR) is then converted to SFR using the conversion 
factor between SFR and $L$(IR) from \cite{ken98}, 
i.e., $\textrm{SFR} { } [M_{\odot}/\textrm{yr}] 
=  4.5 \times 10^{-44} L{\textrm{(IR)}} [\textrm{erg/s}]$. 
Note that our 5$\sigma$ detection limit is 67$\mu$Jy which corresponds
to SFR(IR)$\sim$15 $M_{\odot}$/yr. 
We adopt a 50\% uncertainty in SFR(IR), which is larger than simple
photometric errors, to take into account the reported uncertainty
in the conversion from $\nu L_{\nu , 8\mu \textrm{m}}$ to $L$(IR) 
(e.g. \citealt{cap07}; \citealt{bav08}). 
Using the MIR catalogue constructed by \cite{koy08}, we made 
a catalogue of MIR-detected galaxies within the fields covered
by our MOIRCS observation.  \cite{koy08} constructed two
types of catalogues; one for the 15$\mu$m-resolved members
(single objects) and the other for the 15$\mu$m-unresolved 
members (blended objects).
Likewise, we make such two types of MIR-detected catalogues in this 
paper as well.  We derive star formation rates in IR, SFR(IR),
for the 15$\mu$m-resolved members only, since there are large 
uncertainty in the MIR photometries of the blended sources.

\section{Selection of H$\alpha$\ emitters from cluster members}
\label{sec:selection}

\subsection{Selection technique}
\label{subsec:selection_technique}

If a line emission comes into the NB119 filter, it can be identified as
an object with a flux excess in NB119 compared to the $J$-band flux.
Therefore the $J-$NB119 colour is a good indicator to search for emitters. 
However, we should take into account the fact that the NB119 filter 
($\lambda _c=11885$\AA) is located near the lower-wavelength end of 
$J$-band ($\lambda _c=12600$\AA). Therefore, we cannot simply use 
the $J$-band magnitude as the continuum flux because the continuum 
flux at 1.19~$\mu$m and the $J$-band flux can be different depending 
on the SED slope around that wavelength regime of each galaxy.
Therefore, we estimate the continuum flux level at 1.19$\mu$m of each galaxy
by linearly interpolating the magnitudes in $z'$ and $J$ bands.
This correction substantially reduces the scatter in $J$$-$NB119 
colours around the zero value which is expected for non-emitters 
in NB119 at various redshifts.
We refer to such flux 'corrected' $J$-band magnitude as $J_{\rm{corr}}$, 
and $J_{\rm{corr}}$$-$ NB119 colours of our sample are plotted against
NB119 magnitudes in Fig.~3.  The solid curves show $\pm 3\sigma$ errors in
$J-$NB119 colours.  We define NB119 emitters as those having colour 
excesses of $J_{\rm{corr}}-$NB119 $> 0.3$ and above the upper 
curve at the same time.

 \begin{figure}
   \begin{center}
    \leavevmode
    \vspace{-2.2cm}
    \rotatebox{0}{\includegraphics[width=8.6cm,height=8.6cm]{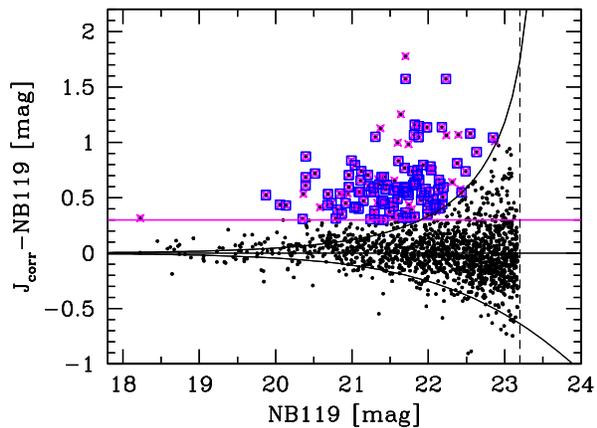}}
    \vspace{-0.7cm}
   \end{center} 
   \caption{The narrow--broad bands colour--magnitude diagram to
define our NB119 emitters.
The $J$-band magnitudes are corrected for a colour-term using
$z'$$-$$J$ colours so that they correspond to the continuum
fluxes at 1.19$\mu$m.
The vertical dashed-line shows 5$\sigma$ limiting magnitudes in NB119 and 
the solid curves indicate $\pm 3\sigma$ excess in 
$J$$-$NB119 colours.  Galaxies above the horizontal solid line ($J$$-$NB119
=0.3) and the upper solid curve are defined as the NB119 emitters 
(shown by crosses), of which the open squares show the H$\alpha$\ emitters 
at $z=0.81$ as defined later (see text and Fig.~\ref{fig:member_selection}). }
\label{fig:emitter_selection}
 \end{figure}
 \begin{figure}
   \begin{center}
    \vspace{-1cm}
    \leavevmode
    \rotatebox{0}{\includegraphics[width=8.5cm,height=8.5cm]{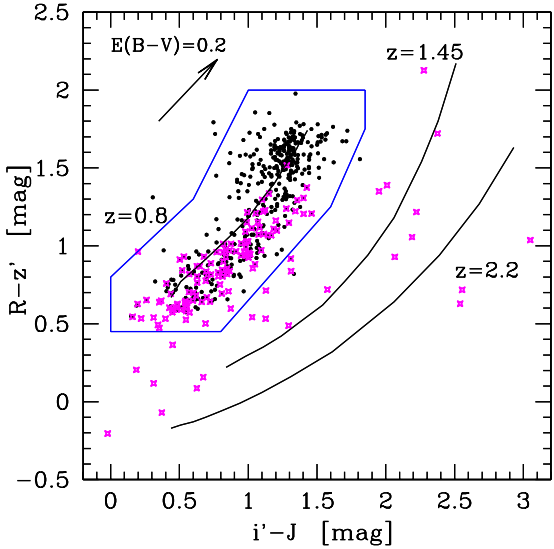}}
   \end{center} 
   \vspace{-0.3cm}
   \caption{The colour--colour diagram to select H$\alpha$\ emitters at
  $z=0.81$. The phot-$z$ selected members (filled circles) and the 
  NB119 emitters (crosses) are plotted. Three solid curves show the
  colour tracks of galaxies at $z=$0.8, 1.45, 2.2, respectively, based 
  on the Kodama et al.~(1999) model. Along the curves, the bulge fraction
  to the total light is changed from 0.0 to 1.0 (blue side to red side) 
  at fixed redshift.
  The NB119 emitters located in the closed box are 
  defined as H$\alpha$\ emitters at $z=0.81$.  The arrow 
  shows a reddening vector of $E(B-V)$$=$0.2, calculated from the
  extinction law of Calzetti et al.~(2000) assuming $R_V$$=$3.1. } 
\label{fig:member_selection}
 \end{figure}

The NB119 emitters thus selected are not necessarily H$\alpha$\ emitters
at $z=0.81$ because other line emitters at other redshifts are also
included in the NB119 emitters. For example, H$\beta$/[OIII] emitters 
at $z=1.4$--$1.5$ and [OII] emitters at $z=2.2$ are the major contaminants.
We eliminate these galaxies using broad-band colours.
We show in Fig.~\ref{fig:member_selection} a colour--colour diagram. 
We plot phot-$z$ selected galaxies (i.e. 0.76$\le z_{\rm{phot}}
\le$0.83) as small filled circles and all the NB119 emitters as 
crosses.  We also show the predicted colour tracks of model galaxies 
at $z=0.8, 1.45, 2.2$, respectively, along which the contribution of 
bulge component is changed from 0.0 to 1.0 at each redshift \citep{kod99}.
We can see that some emitters have very different colours from 
the model track of $z=0.8$ galaxies or the distribution of the phot-$z$
selected galaxies. These galaxies are likely to be contaminants.
We set a boundary shown by a closed box in Fig.~\ref{fig:member_selection},
and define the H$\alpha$\ emitters at $z=0.81$ as the NB119 emitters 
that satisfy these criteria. We may miss a few real members near 
the boundary, but their effect to our conclusion is negligible.

We summarize our definitions of cluster members in the following.
Firstly, we select all galaxies with $0.76 \le z_{\rm{phot}} \le 0.83$ 
as members.
This criterion recovers more than 90 \% of spectroscopically confirmed
members listed in \cite{gio99} (see fig.~2 in \citealt{koy07}). 
The same criterion is actually used in \cite{koy07} and \cite{koy08}. 
The number of these phot-$z$ selected members is 447. 
Secondly, we include all the H$\alpha$\ emitters at $z=0.81$ selected 
above.  We have 114 H$\alpha$\ emitters in total. Note that 32 of them
have been missed out by our phot-$z$ selection, although many of them have
$z_{\textrm{phot}}\sim$ 0.70--0.75, only slightly lower than our
phot-$z$ selection criterion. 
This is inevitable since the phot-$z$ tends to give lower
redshift estimation for star-forming galaxies due to the
lack of prominent features in their SED,
while redshift estimation for passively evolving galaxies are
relatively more accurate (e.g. \citealt{kod99}).
This means that the combined approach that we take in this paper, namely
the phot-$z$ selection and the narrow-band emitter survey, is quite effective
to increase the completeness of true cluster members.

\subsection{Local density measurements}
\label{subsec:emitter_selection}

One of the main purposes of this paper is to examine environmental
dependence of H$\alpha$-derived star formation activity at $z=0.8$.
To quantify the environment, we estimate the local density of each 
member galaxy using its five nearest neighbours.  We already defined
the local density ($\Sigma _5$ in Mpc$^{-2}$ unit) in \cite{koy08}, 
but we recalculated
it since we added 32 new members in this paper as shown in the previous
section.  When we calculate $\Sigma _5$, 
we use all the member galaxies used in \cite{koy08} (i.e. all the 
$\sim 2700$ galaxies throughout the Suprime-Cam field in the optical 
catalogue with $0.76 \le z_{\textrm{phot}} \le 0.83$ to the depth 
of $z'$=24.9) and the 32 newly selected members.
Therefore, it makes almost no change for the estimation 
of $\Sigma_{5}$ from that of \cite{koy08}, but the main purpose 
here is to define environment with $\Sigma _5$ in the same way
for the newly added members. 

Following the definition in \cite{koy08}, we classify galaxies into three
environmental bins, namely, 'low-density', 'medium-density' and 'high-density' 
regions which correspond to 
$\log\Sigma _5 < 1.65$, $1.65 \le \log\Sigma _5 < 2.15$ and 
$\log \Sigma _5 \ge 2.15$, respectively. 
The high-density region corresponds to the cluster core. The medium-density
region corresponds to cluster outskirts, groups, and filaments.
The low-density region corresponds to outer fields.   
We note that as shown in \cite{koy08} the optical colour distribution 
starts to change dramatically from blue to red in the medium-density 
environment thus defined (see also \citealt{kod01}; \citealt{tan05}).

\section{H$\alpha$-derived star-formation rate and specific star formation rate}
\label{sec:Ha_measurements}

\subsection{Derivation of SFR(H$\alpha$) and SSFR(H$\alpha$)}

 \begin{figure}
   \begin{center}
    \leavevmode
    \vspace{-2cm}
    \rotatebox{0}{\includegraphics[width=8.5cm,height=8.5cm]{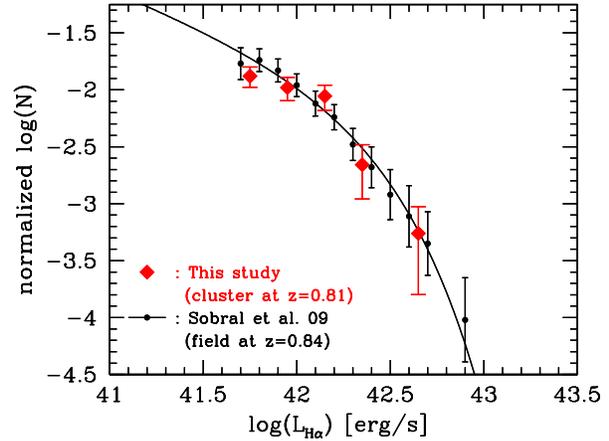}}
   \end{center} 
   \vspace{-0.3cm}
   \caption{ H$\alpha$\ luminosity function of the RXJ1716 cluster
  (filled diamonds with purely poissonian error bars). The normalization
  of the cluster LF is arbitrarily scaled to compare with the field 
  H$\alpha$\ luminosity function of Sobral et al.~(2009) (filled circles
  and their best-fitted LF as shown by the solid curve).  }
\label{fig:Ha_LF}
 \end{figure}
  \begin{figure*}
   \begin{center}
    \vspace{-0.2cm}
    \leavevmode
    \epsfxsize 0.47\hsize
    \epsfbox{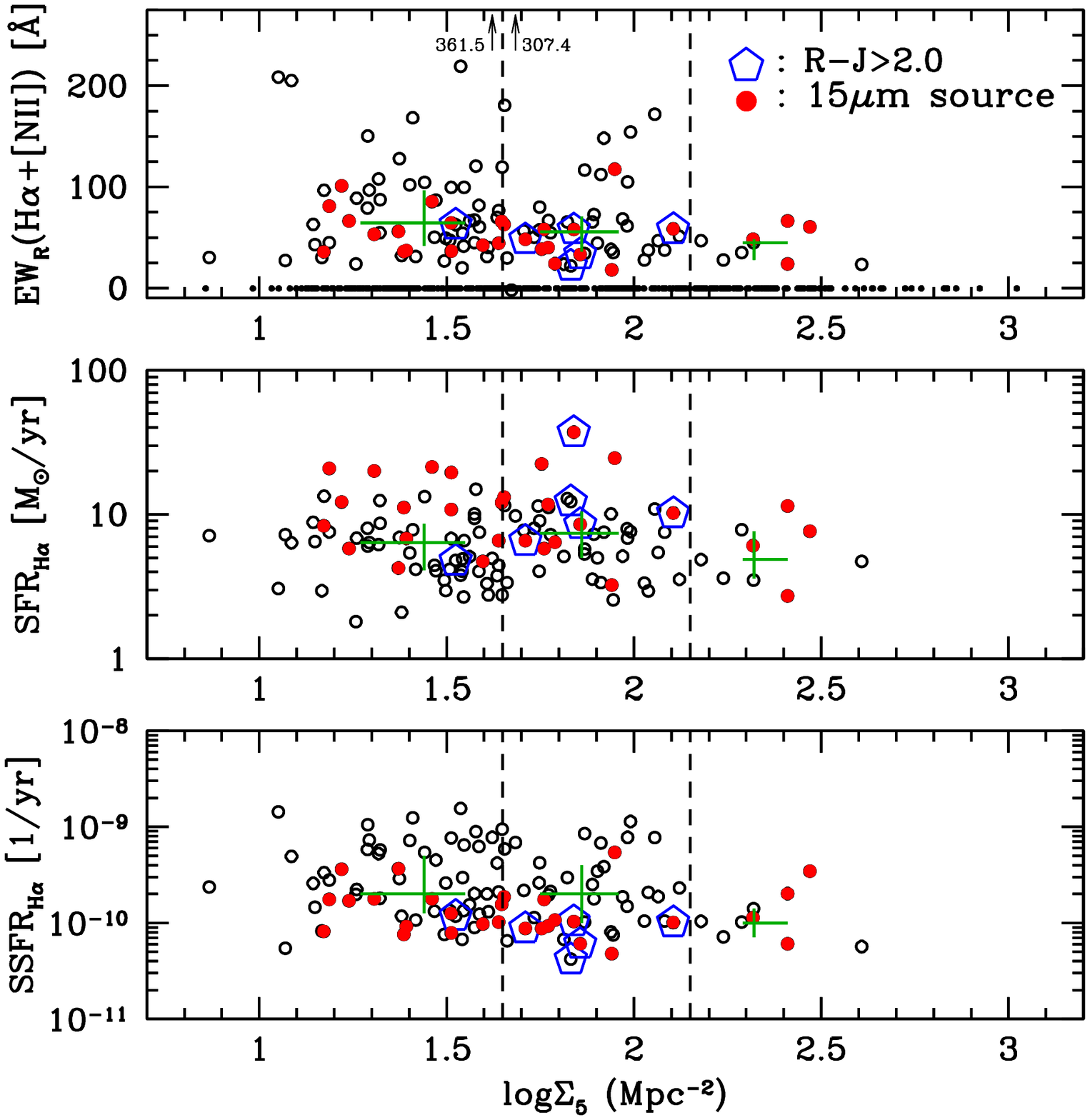}
    \epsfxsize 0.47\hsize
    \epsfbox{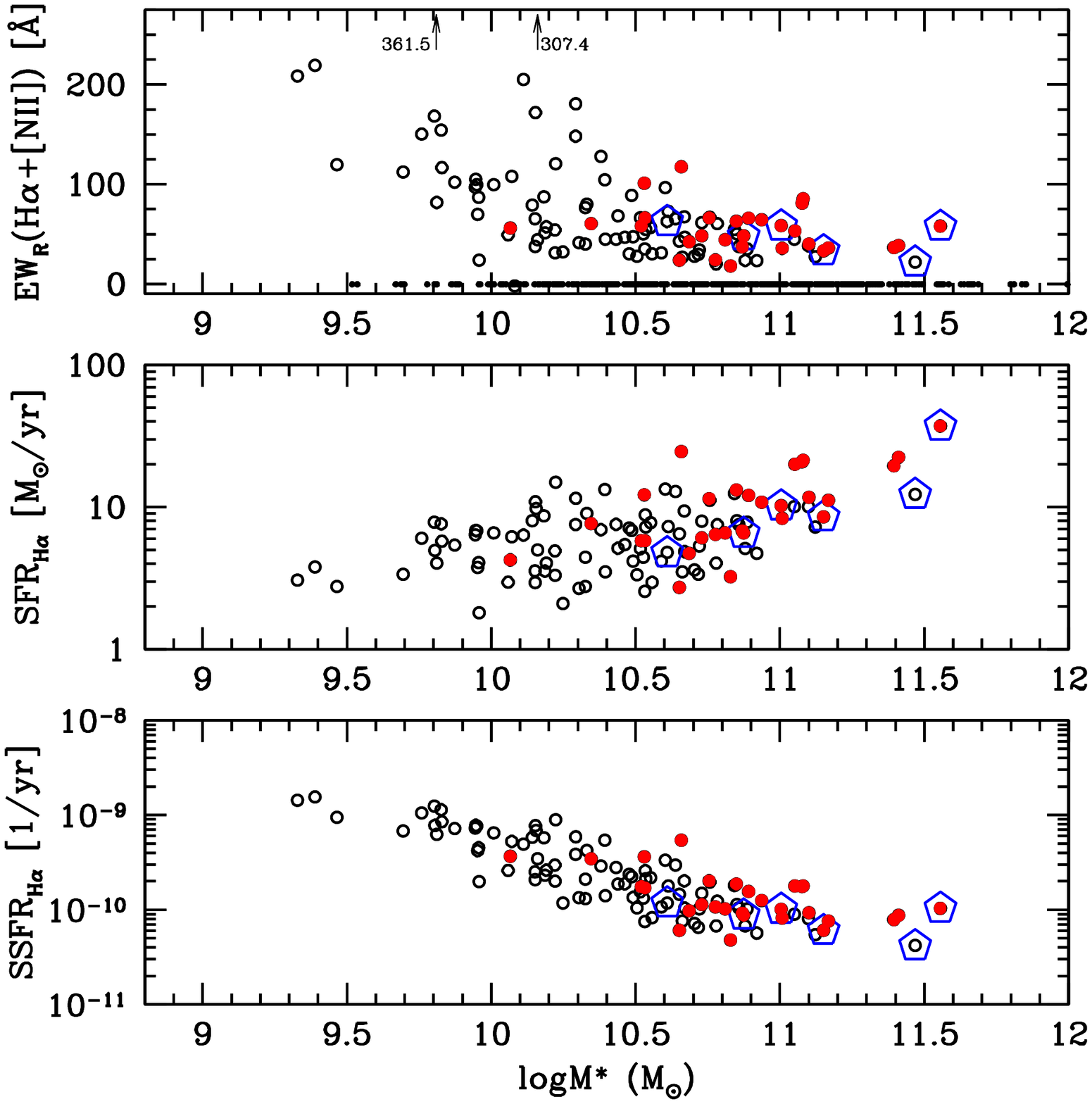}
   \end{center} 
  \vspace{0cm}
   \caption{The physical quantities of the H$\alpha$\ emitters,
EW(H$\alpha$+[NII]), SFR(H$\alpha$) and SSFR(H$\alpha$),
are plotted as a function of the local density ({\it left}) and stellar mass 
({\it right}). The open circles show all the H$\alpha$ emitters, the 
filled circles represent the MIR galaxies and the open pentagons 
indicate the red H$\alpha$\ emitters with $R-J>2.0$.
In SFR and SSFR, we have applied a correction for 1 mag extinction.
In the top panels, the non-H$\alpha$\ emitters are also plotted at EW$=$0,
while only the H$\alpha$\ emitters are plotted in the middle and the 
lower panels. The two small arrows with numbers in the top panels
indicate the galaxies with high EW(H$\alpha$+[NII]) located above the
top boundary. In the left panels, we show median values as crosses 
in each environment with error bars showing 25--75\% distribution
of each sample. 
The vertical dashed lines at $\log \Sigma _5 =$1.65 and 2.15 in the 
left panels indicate the dividing lines between the low-, medium- 
and high-density environments as defined in Koyama et al.~(2008).
 }
\label{fig:SFR_vs_env}
 \end{figure*}

In this subsection, we calculate some basic physical quantities to characterize
the star formation activity, such as star-formation rates (SFRs) and
specific star-formation rates (SSFRs), for all the H$\alpha$\ emitters.
Firstly, we derive H$\alpha$\ line flux, continuum flux and 
rest-frame equivalent width by the following equations:
\begin{equation}
F_{\rm{H\alpha + [NII]}}=\Delta_{\textrm{NB}}\frac{ f_{\textrm{NB}} -
 f_J }{1-\Delta_{\textrm{NB}}/\Delta_{J}}
\end{equation}
\begin{equation}
f_c = \frac{f_J - f_{\rm{NB}}(\Delta_{\rm{NB}}/\Delta_{J})}{1-\Delta_{\rm{NB}}/\Delta_{J}}
\end{equation}
\begin{equation}
\textrm{EW}_R({\rm{H\alpha + [NII]}}) = (1+z)^{-1} \frac{F_{\rm{H\alpha}}}{f_c} 
\end{equation}
where $\Delta_J$ and $\Delta_{\rm{NB}}$ are the widths in wavelength 
of the $J$ and NB119 filters, $f_J$ is the flux density at 1.19~$\mu$m 
derived from the $J_\textrm{corr}$, and $f_{\rm{NB}}$ is the flux 
density for NB119. We then multiply 4$\pi$$d_L^2$ to $F_{\rm{H\alpha +[NII]}}$ 
to derive the luminosity $L$(H$\alpha$+ [NII]), where $d_L$ is the luminosity 
distance $5.10 \times 10^3$Mpc at $z=0.81$. Finally, we compute the 
H$\alpha$-based star formation rates, SFR(H$\alpha$), using the relation 
of \cite{ken98}, SFR(H$\alpha$)[$M_{\odot}$/yr]$=7.9\times
10^{-42} L_{\rm{H\alpha}}$[erg/s].  
We note that our selection criteria of H$\alpha$\ emitters
shown in Section~3.1 correspond to EW$_{\textrm{R}}$(H$\alpha$+[NII])
$\gsim$30\AA {} and SFR(H$\alpha$)$\gsim$1$M_{\odot}$/yr before 
extinction correction.  We correct for 30\%
[NII] line contribution (e.g. \citealt{tre99}) and consider 1 mag 
extinction of H$\alpha$\ (\citealt{ken94}).  
We should note that these assumptions adopted here are somewhat 
uncertain and can only be applied to discuss statistical behaviours.
In particular, the amount of extinction can be much larger 
for very actively star-forming galaxies (e.g. \citealt{pog00})
and this will be discussed in Section~7 of this paper.  
We also note that we do not correct for the contamination of
active galactic nuclei (AGNs) in our H$\alpha$\ emitters sample.
In reality, some H$\alpha$\ emission lines may be originated from AGNs
rather than star-forming H{\sc ii} regions.
However, it is reported that the fraction of AGNs in distant
clusters is very small (only $\sim 1$\%) from optical spectroscopic
surveys (e.g.\ \citealt{dre99}).
The fraction of X-ray selected AGNs could be higher by a factor of
$\sim 5$ than that of optically selected ones (e.g. \citealt{mar02}).
But in any case, AGN contamination is not a major concern and has
little effect on our statistical discussion on the H$\alpha$\ emitters
(see also \citealt{gar09}).

One of the best indicators to quantify the status of galaxy activity
is the specific star-formation rate (SSFR), which is SFR divided by 
stellar mass of galaxies (i.e. SFR per unit stellar mass).
We derive stellar mass ($M_*$) of galaxies in the following way.
We use the relation between $M_*$/$L_{J(\textrm{obs})}$ of galaxies 
and $R-J$ at $z=0.8$ for the model galaxies with different 
contribution of bulge component from \cite{kod99}.
Note that the stellar mass in the model is re-scaled using the
Salpeter IMF for consistency with the derivation of SFR.
Based on the observed $R-J$ colour of each galaxy, we derive 
$M_*$/$L_{J(\textrm{obs})}$ which is then multiplied to the observed 
$J$-band luminosity to get $M_*$. SSFR is derived by dividing SFR by 
$M_*$.

\subsection{H$\alpha$\ luminosity function}
\label{subsec:Ha_LF}
In Fig.~\ref{fig:Ha_LF}, we show the H$\alpha$\ luminosity function (LF) 
using H$\alpha$\ emitters within 2.5 Mpc from the cluster centre. 
We here slightly modify the selection criteria of H$\alpha$\ emitters
to make a fair comparison with \cite{sob09}, who constructed 
a field H$\alpha$\ LF at $z\sim 0.8$.  
Assuming that all H$\alpha$\ emitters are associated with the cluster,
we do not apply field subtraction. 
Our cluster LF is arbitrarily scaled to match the field H$\alpha$\ LF 
from \cite{sob09} (filled circles with their best-fitted LF). 
Therefore, we can only compare the shape of the cluster and 
field H$\alpha$\ LFs. It is apparent that there is no significant 
difference in the shape of the H$\alpha$\ LF.
This is similar to the result from \cite{kod04}. They showed 
that the H$\alpha$\ luminosity function shows little sensitivity
to local density, using their very wide H$\alpha$\ imaging 
around CL0024 cluster at $z=0.39$.  In the local Universe, 
this is also true (\citealt{bal04}). \cite{bal04} limit their
sample for star-forming galaxies and showed that the strength 
of H$\alpha$\ lines does not depend on environment. 
Thus, although the fraction of star-forming galaxies may change 
as a function of environment, star-forming activity of star-forming
galaxies seems not strongly dependent on environment. 
Our new H$\alpha$\ study for the $z=0.81$ cluster extends this idea 
up to $z\sim 0.8$. As will be discussed in Section~7, 
there are some exceptionally active galaxies in the cluster 
surrounding environment. However,  
it would be difficult to find such excess of activity in this
H$\alpha$\ LF.  At least, our comparison suggests that we will need 
much larger sample to detect the environmental difference
of star-forming activity {\it among} the star-forming galaxies,
if any.

\subsection{Dependence on environment and stellar mass}

In Fig.~\ref{fig:SFR_vs_env}, we plot
EW$_{\textrm{R}}$(H$\alpha$+[NII]), SFR(H$\alpha$) and SSFR(H$\alpha$) 
of H$\alpha$\ emitters as a function of local density and stellar mass. 
It is apparent that very strong H$\alpha$\ emitters with 
EW$\gsim$100\AA{} are seen only in the medium- and the low-density 
environments, and no such galaxy is found in the high-density 
environment (see the top-left panel of Fig.\ 6), although this may be 
due to small number statistics of the H$\alpha$ emitters in the 
high-density region. Other than that, we see no significant 
environmental trends in these quantities of the H$\alpha$\ emitters.
Rather, they show modest correlations with 
stellar mass of galaxies (right panels). Strong H$\alpha$\ emitters with
EW$\gsim$100\AA {} are found only among low mass galaxies with 
$\log M_* \lsim 10.4$. More massive galaxies have higher SFRs but 
lower SSFRs than less massive galaxies. 

\begin{figure}
   \begin{center}
    \leavevmode
    \rotatebox{0}{\includegraphics[width=8.5cm,height=8.5cm]{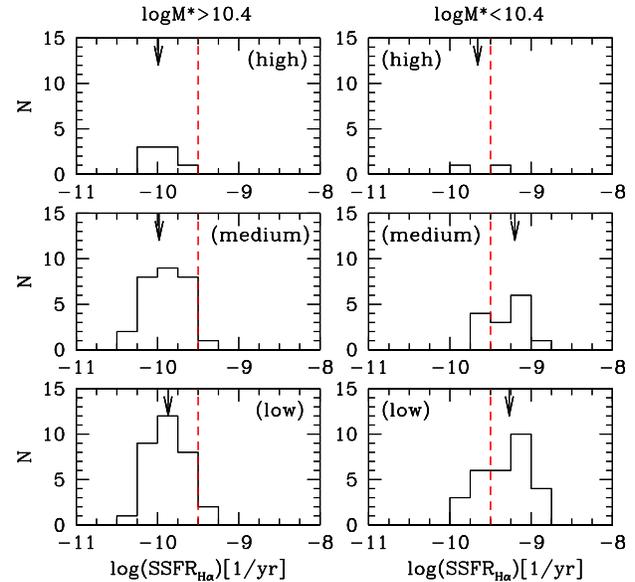}}
   \end{center} 
   \vspace{-0.4cm}
   \caption{The distributions of SSFR for high-mass 
  ($\log M_*/M_{\odot}$$>$10.4) and low-mass 
  ($\log M_*/M_{\odot}$$<$10.4) galaxies in each environmental bin for all 
the H$\alpha$\ emitters.  The arrows indicate the median value of SSFR 
in each panel. The vertical dashed lines at 
SSFR$=$$-$9.5 show the dividing line between the high and low SSFR
galaxies for an eye-guide. }
\label{fig:SSFR_histogram}
 \end{figure}

We show in Fig.~\ref{fig:SSFR_histogram} the distribution of SSFR(H$\alpha$)
for each environment, dividing our H$\alpha$\ emitter samples into 
low-mass ($\log M_*/M_{\odot}$$<$10.4) and high-mass 
($\log M_*/M_{\odot}$$>$10.4) 
galaxies. We find that high-mass galaxies tend to have low
SSFR (SSFR$\lsim$$-9.5$) compared to low-mass galaxies in all
environment. The lack of low SSFR galaxies in low-mass galaxies 
might be partly due to the selection effect, since the selection
criteria of NB emitters is more strict for fainter galaxies as can
be seen in Fig.~\ref{fig:emitter_selection}. 
However, the lack of high mass galaxies with high SSFR 
should be reliable.   
We show the median value of SSFR(H$\alpha$) in each panel as an arrow. 
We find that the median values do not strongly correlated with 
environment if we fix the stellar mass of galaxies, although the
number of low-mass H$\alpha$\ emitters in high-density environment
is really small. 
We should note that this suggestion is valid only when 
we limit the sample to the H$\alpha$\ emitters, and we should keep 
in mind that there are many H$\alpha$-undetected galaxies in 
particular in higher-density environments (see Section~5).

We plot the MIR-detected galaxies as filled symbols 
in Fig.~\ref{fig:SFR_vs_env}.
We can notice that the MIR-detected galaxies tend to have 
low EW(H$\alpha$+[NII]) and low SSFR(H$\alpha$) in spite of their
high star formation rates in MIR (SFR(IR)$\gsim$15$M_{\odot}$/yr).  
One may claim that MIR-detected galaxies tend to be biased to
more massive galaxies with larger SFR. However, interestingly, there
is no significant difference in SFR(H$\alpha$) between 
MIR-detected and MIR-undetected galaxies (see middle-right panel
in Fig.~\ref{fig:SFR_vs_env}). There is a trend that
massive H$\alpha$\ emitters tend to be detected in MIR, indicating
that they tend to be more dusty, hence their SFRs (and SSFRs) are
likely to be underestimated even with H$\alpha$.
This suggests that we need not only H$\alpha$\ information
but also MIR information to fully understand the star formation
history of massive galaxies in the distant Universe. 
In the remaining of this paper, taking the great advantage of wide-field
coverage of our H$\alpha$\ and MIR data, we attempt to show how large
amount of star-formation is obscured and hidden in the previous optical
surveys and how the obscuration is related to environment at $z\sim 0.8$
for the first time.

\section{Mapping out star-formation activity in and around the cluster}

\subsection{Spatial distribution}
\label{subsec:spatial_distribution}

 \begin{figure*}
   \begin{center}
    \leavevmode
    \epsfxsize 0.48\hsize
    \epsfbox{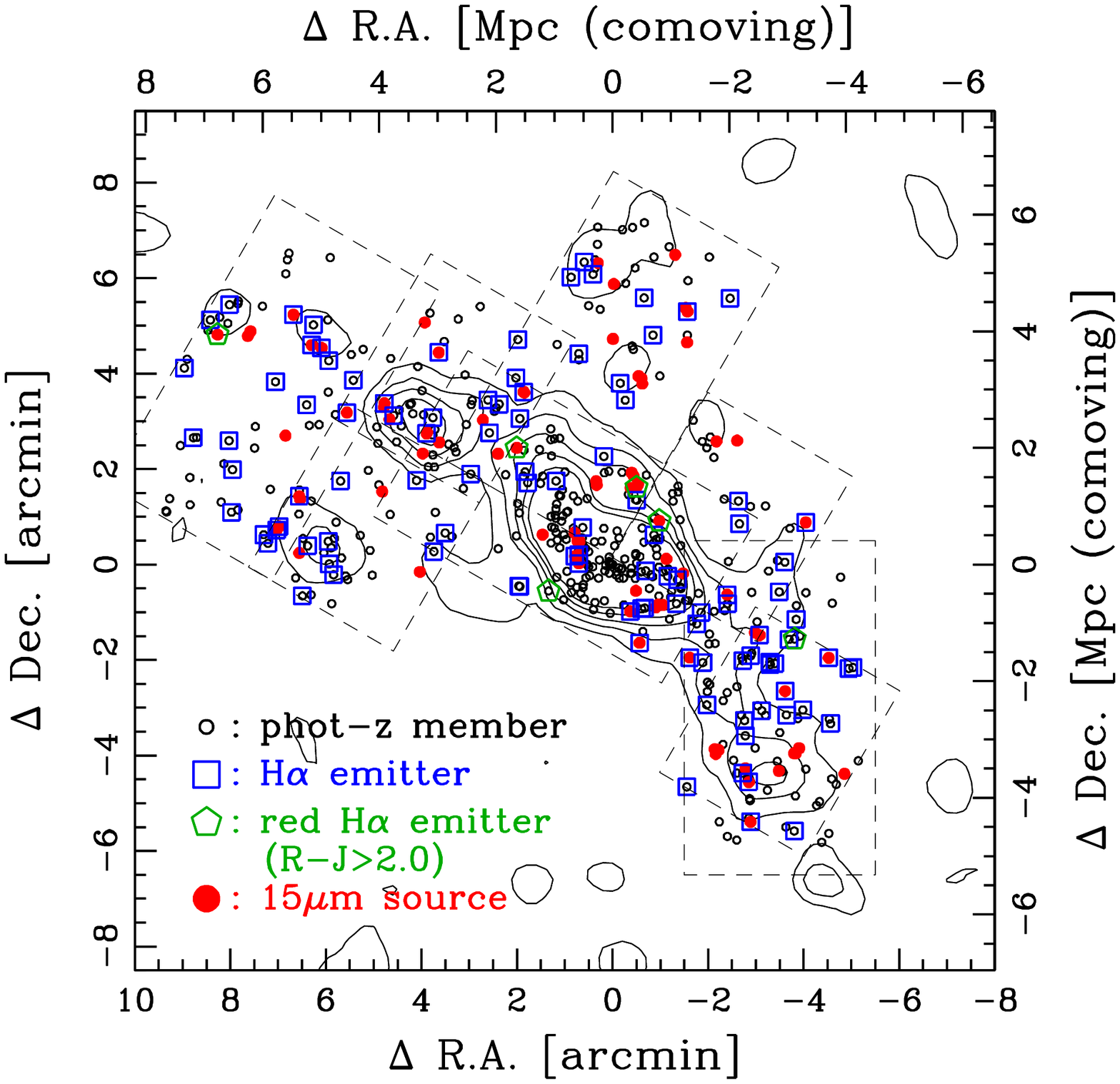}
    \epsfxsize 0.48\hsize
    \epsfbox{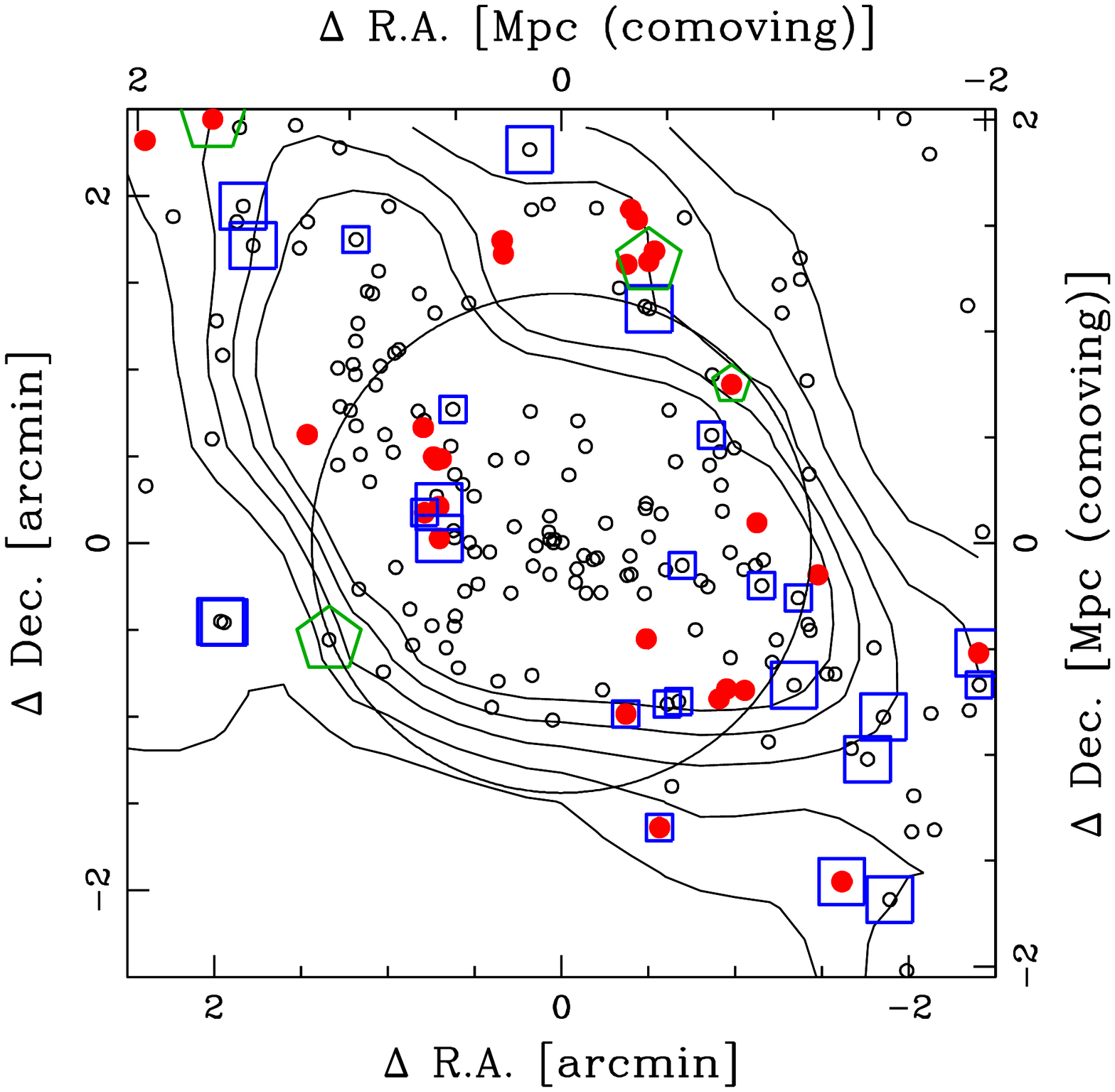}
   \end{center} 
   \vspace{-0.4cm}
   \caption{ ({\it left}): The 2-D distributions of all the members
 (open circles), the MIR-detected galaxies (filled circles),
  the blue H$\alpha$\ emitters with $R-J$$<$2.0 (open squares), 
  and the red H$\alpha$\ emitters with $R-J$$>$2.0 (open pentagons). 
  The dashed-line boxes indicate our MOIRCS FoVs.  We plot the member 
  galaxies detected in our MOIRCS FoVs only.  The contours are the 
  same as in Fig.~2. 
  ({\it right}): A close-up view of the cluster central region.
  The meanings of the symbols are the same as in the left panel,
  except that the large and the small open symbols indicate the strong
 (EW$>$ 50\AA) and weak (EW$<$ 50\AA) H$\alpha$\ emitters, respectively.
  The solid-line circle shows the radius of 
  $0.5 \times R_{\textrm{200}}$ from the cluster centre
  (see Section~8). }
\label{fig:map}
 \end{figure*}

We show in the left panel of Fig.~\ref{fig:map} the spatial distribution of 
cluster member galaxies (small open circles), blue H$\alpha$\ emitters 
with $R-J<$2.0 (open squares) and red H$\alpha$\ emitters 
with $R-J>$2.0 (open pentagons). We only plot the galaxies in the 
area observed by the NB filter. We can see that H$\alpha$\ emitters well 
trace the filamentary large-scale structures.  
This is a strong evidence for that most of the small groups identified 
by phot-$z$ are physically associated with the cluster and not just a 
concentration of fore-/background galaxies.
Note that we may be missing some H$\alpha$\ emitters especially
in the outskirt of the clusters, where relative velocities of
galaxies with respect to the cluster core would be larger,
although the width of the NB119 filter neatly covers the radial velocity
range of $\pm$ $\sim$1500~km s$^{-1}$ (see Fig.~1).
A 5$' \times$5$'$ close-up view is shown in the
right panel of Fig.~\ref{fig:map}. We also plot the position of 
MIR-detected galaxies (including resolved and unresolved objects) 
as filled circles. 
It can be easily noticed that the distribution of H$\alpha$\ emitters 
is quite similar to that of MIR-detected galaxies.  We reported 
the avoidance of MIR-detected galaxies in the cluster central region
in \cite{koy08}, and this study strengthens the conclusion that the
star-forming activity is very low in the cluster core of this cluster. 

On the other hand, it is interesting to note that the  
H$\alpha$\ emitters and MIR-detected galaxies do not always 
directly overlap each other. 
Since our H$\alpha$\ survey can go deeper in terms 
of SFR ($\sim 1$ $M_\odot$/yr without extinction correction) 
compared to MIR observation ($\sim$ 15 $M_{\odot}$/yr), it is 
natural to expect that some moderately 
(or weakly) star-forming galaxies are detected only in H$\alpha$. 
In fact, we can see many such galaxies in Fig.~\ref{fig:map}. 
However, there are also some galaxies which are detected in MIR but 
not in H$\alpha$. These type of galaxies are difficult to interpret.
It may be possible that these galaxies are very dusty and a large amount
of UV/optical light is hidden by dust, so that we cannot detect their
H$\alpha$\ emission lines. 
It is also possible that some MIR galaxies are located outside of 
the redshift range of the H$\alpha$\ emitters that is covered by 
the NB119 filter. Some of them can still be members (see Fig.\ 1), 
but some could be non-member contaminants due to the limitation 
of phot-$z$ based membership determination of the MIR galaxies.
Spectroscopic confirmation of their membership is needed to 
discuss this interesting population further.

\subsection{H$\alpha$\ fraction}
\label{subsec:Ha_fraction}

\begin{figure}
   \begin{center}
    \leavevmode
    \vspace{-1cm}
    \rotatebox{0}{\includegraphics[width=8.6cm,height=8.6cm]{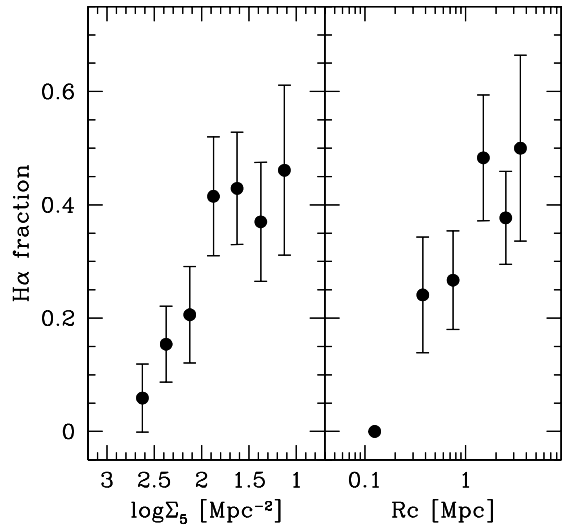}}
   \end{center} 
  \vspace{-0.5cm}
   \caption{The fraction of the H$\alpha$\ emitters as a function of
local density ({\it left}) and as a function of cluster-centric distance
({\it right}). The error-bars show Poisson errors. }
\label{fig:Ha_fraction}
 \end{figure}

As shown in the previous section, H$\alpha$\ emitters avoid
cluster central region and/or high-density region. We quantify it
by calculating the fraction of H$\alpha$\ emitter as functions of local 
density and cluster centric radius (Fig.~\ref{fig:Ha_fraction}). 
We use galaxies with NB119$<$22.0 where the selection of 
H$\alpha$\ emitters is complete down to $J-$NB119$=$0.3 
(see Fig.~\ref{fig:emitter_selection}). 
It can be seen that the H$\alpha$\ fraction is strongly dependent
on local density and cluster centric radius, in the sense that
the fraction is lower in the higher-density environment.
Thus, we conclude that there is little star-formation activity
in the core of the RXJ1716 cluster.
Such global trend that the fraction of H$\alpha$ emitters decreases
towards higher density regions (Fig. 9) is largely due
to the fact that the fraction of passive galaxies increases towards
higher density regions.  In fact, if we limit the sample only to blue
galaxies with $R-J<2.0$, the fraction of H$\alpha$ emitters does no longer
strongly correlate with local density or cluster centric radius,
and is almost constant at $\sim$60\%. This is consistent with the results
obtained in Section 4.2 and 4.3 that the observed H$\alpha$ line strength
in star-forming galaxies does not strongly correlate with environment.

\cite{fin05} conducted similar H$\alpha$\ emitter survey for the 
central part of three $z\sim 0.7$ clusters. 
They showed that two of the three clusters have few 
H$\alpha$\ emitters in the central region.
However, they found many H$\alpha$\ emitters in the core region
of the other cluster. \cite{hay09} show that there are many [OII] emitters
even in the cluster core at $z=1.46$, and that the fraction of [OII] 
emitters is still high in the core within $R_{\rm c}$$<$0.25 Mpc from
the cluster centre. For the RXJ1716 cluster, we find no H$\alpha$\ 
emitters at $R_{\rm c}$$<$0.25 Mpc as shown in the right-panel 
of Fig.~\ref{fig:Ha_fraction}, although there are 31 member galaxies 
in $R_c$$<$0.25 Mpc without significant detection in H$\alpha$. 
Therefore, for the RXJ1716 cluster, the termination of star-forming 
activity in the cluster core has been completed before $z=0.8$.
These facts suggest that the epoch of the termination of star-forming 
activity in the cluster core may differ from cluster to cluster, 
probably depending on the cluster mass and/or maturity of clusters.
We need a much larger sample of distant clusters to draw a general 
picture of formation and evolution of the cluster cores.

 \begin{figure*}
   \begin{center}
    \leavevmode
    \epsfxsize 0.48\hsize
    \epsfbox{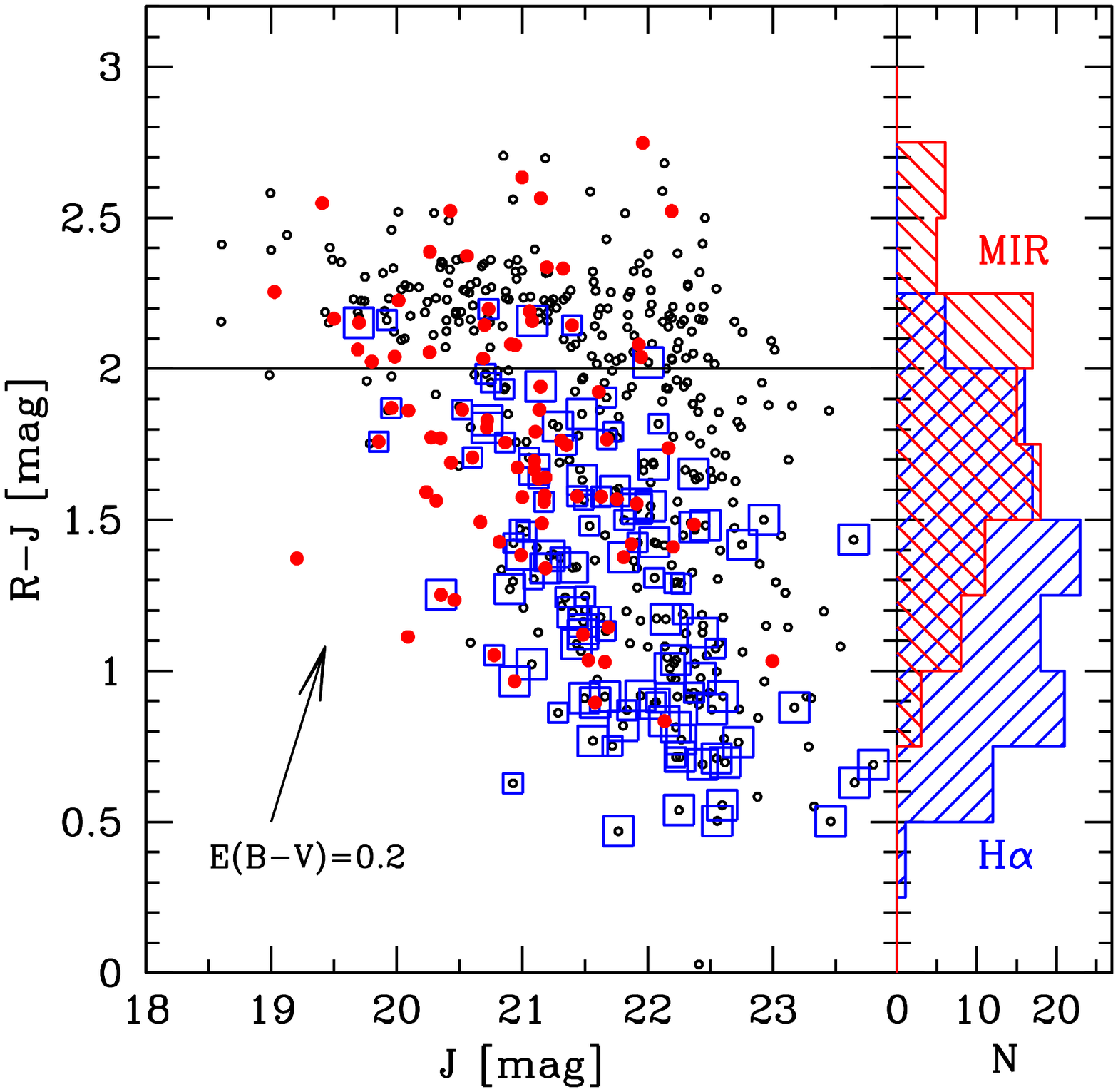}
    \epsfxsize 0.48\hsize
    \epsfbox{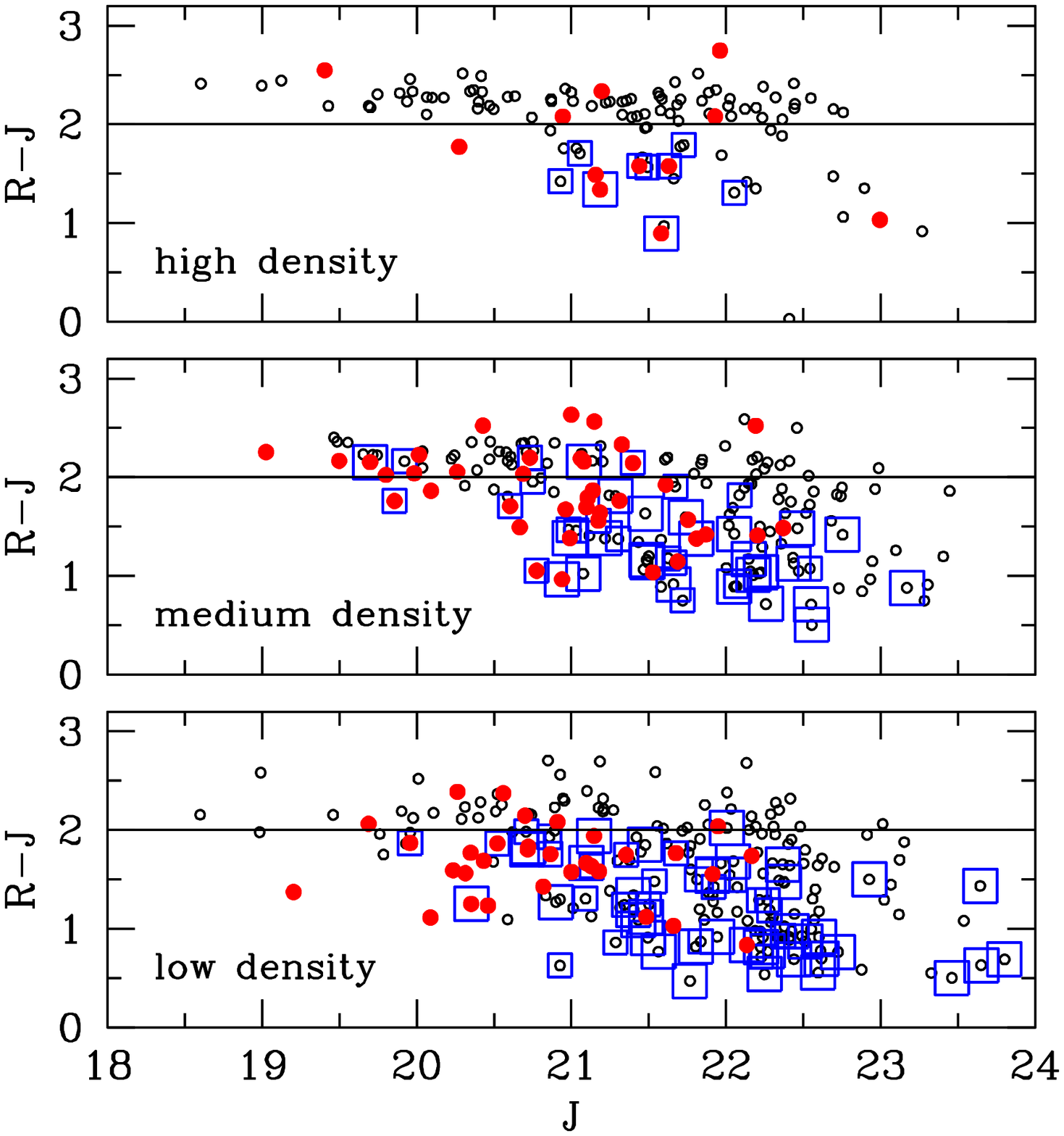}
   \end{center} 
\caption{({\it left}): The colour--magnitude diagram for all the 
observed fields. The open circles show all the member galaxies.  
The small and large open squares 
indicate the weak (EW$<$50\AA) and strong (EW$>$50\AA) H$\alpha$\ emitters,
respectively.  The filled circles indicate the MIR galaxies. 
The histograms show the colour distributions of the H$\alpha$\ emitters 
and the MIR galaxies as indicated. The arrow shows a reddening vector
that corresponds to $E(B-V)=0.2$.  
({\it right}): The same plot for each environment (low-, medium- and 
high-density regions). The meanings of the symbols are the same as in
the left panel.} 
\label{fig:colmag}
\end{figure*}

\section{Properties of H$\alpha$\ emitters and MIR galaxies}
\label{sec:nature_of_Ha_emitters}

\subsection{Colour--magnitude diagram}
\label{subsec:colour_magnitude_diagram}

In Fig.~\ref{fig:colmag}, we show ($R-J$) v.s. $J$ 
colour--magnitude diagram using all the member galaxies. 
H$\alpha$\ emitters are marked with open squares again. 
The small and large sizes of squares indicate EW(H$\alpha$+[NII])$<$50\AA{}
and EW(H$\alpha$+[NII])$>$50\AA , respectively. 
We clearly notice that H$\alpha$\ emitters are mainly
blue galaxies and that bluer galaxies tend to have stronger 
H$\alpha$\ emission in an average sense.
There are also some H$\alpha$\ emitters on the red sequence. These 
populations are very rare, but interestingly, we find that such 
H$\alpha$\ emitters with red colours are preferentially found 
in the cluster outskirts (see below). 
On the other hand, MIR galaxies (shown as filled circles in 
Fig.~\ref{fig:colmag}) are more commonly seen on the red sequence. 
This is clearly seen in the histograms shown in Fig.~\ref{fig:colmag}.
The H$\alpha$\ emitters are bluer and the MIR galaxies are redder.
This may be because the MIR galaxies are dusty and reddened
in spite of their strong star formation (see also \citealt{koy08}).

We also make the same plot for each environment (i.e. 
for low-, medium- and high-density regions) in the 
right-panels of Fig.~\ref{fig:colmag}. Interestingly, the red sequence
appears quite differently with environment. 
Most of the red galaxies in the high-density and low-density 
environment show little on-going star-forming activity. In contrast, 
a large number of red galaxies in the medium-density environment 
(i.e. in the cluster outskirts, groups and filaments) are detected
in H$\alpha$\ and/or MIR. We detect 6 red H$\alpha$\ emitters
in total, of which 5 are found in the medium-density environment. 
We have already reported in \cite{koy08} a high fraction of 
MIR-detected dusty red galaxies in the medium-density environment.
The above result on the H$\alpha$\ emitters supports this idea. 
\cite{sai08} showed that the number of MIR galaxies on the red sequence
increases towards distant clusters up to $z\sim 0.8$. 
Our current result suggests that the MIR view of the red sequence
is also dependent on environment at $z\sim 0.8$.

\subsection{Colour--colour plot}
\label{subsec:colour_colour_plot}

\begin{figure}
   \begin{center}
    \leavevmode
    \vspace{-0.5cm}
    \begin{center}
    \leavevmode
    \rotatebox{0}{\includegraphics[width=8.6cm,height=8.6cm]{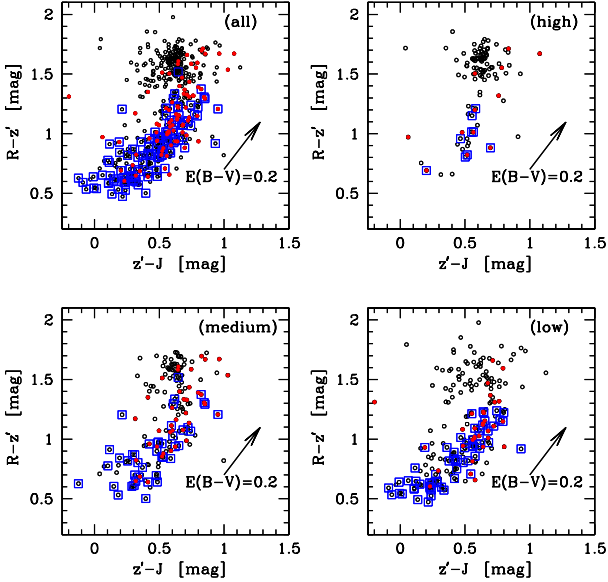}}
    \end{center} 
   \end{center} 
   \vspace{-0.3cm}
   \caption{The colour--colour diagram ($R-z'$ vs.\ $z'-J$) for all the
  observed fields (top-left panel), the high-density region (top-right), 
  the medium-density region (bottom-left), and the low-density region 
  (bottom-right), respectively.  The small open circles represent all
  the member galaxies.  The open squares and the
  filled circles indicate the H$\alpha$\ emitters and the MIR galaxies, 
  respectively. The arrows indicate reddening vectors corresponding to
  $E(B-V)$$=$0.2.}
\label{fig:color_color}
 \end{figure}

We show in Fig.~\ref{fig:color_color} the colour--colour diagrams
($R$$-$$z'$ v.s. $z'$$-$$J$).
Using the three bands that neatly straddle the rest-frame 4000\AA\ break,
we can effectively distinguish dusty red galaxies from passively evolving
red galaxies.  This method is similar to the one used to distinguish
between dusty and passive galaxies for the sample of the extremely
red objects (EROs) at $z\sim1.5$ (e.g. \citealt{poz00}).

In the top-left panel of Fig.~\ref{fig:color_color}, we notice that the
star-forming galaxies (i.e. H$\alpha$\ emitters and MIR galaxies)
are distributed in an elongated region along the direction of 
the reddening vector (shown by the arrow).
In contrast, non-star-forming galaxies (small dots) are concentrated
at around $R-z' \sim 1.6$ and $z'-J \sim 0.6$.
Thus, we can separate between passive galaxies and star-forming galaxies
on this diagram.
Many of the MIR galaxies are red in a single colour in Fig.~\ref{fig:colmag}, 
but with an appropriate colour combination here, the passive galaxies and the
dusty star-forming galaxies occupy different areas on the diagram and can
be well separated.

In Fig.~\ref{fig:color_color}, we also make the same plot for each 
environment. As is also seen in Fig.~\ref{fig:colmag}, H$\alpha$\ 
emitters are blue and MIR galaxies are redder.
Most of the red star-forming galaxies are detected only in MIR 
and even not in H$\alpha$.  Such red MIR galaxies are frequently seen
in the medium-density environment. We also find that very blue galaxies 
are detected only in H$\alpha$\ but not in MIR (as seen in 
Fig.~\ref{fig:colmag}).
This may be because bluer galaxies tend to be lower-mass galaxies 
(see Fig.~\ref{fig:colmag}) and to have lower SFR 
(see Fig.~\ref{fig:SFR_vs_env}), so that we could not detect 
many of these low-mass star-forming galaxies 
by our MIR observation due to the MIR flux limit.

\subsection{Colour--density plot}
\label{subsec:colour_density}

\begin{figure}
   \begin{center}
    \vspace{-0.3cm}
    \leavevmode
     \rotatebox{0}{\includegraphics[width=8.6cm,height=8.6cm]{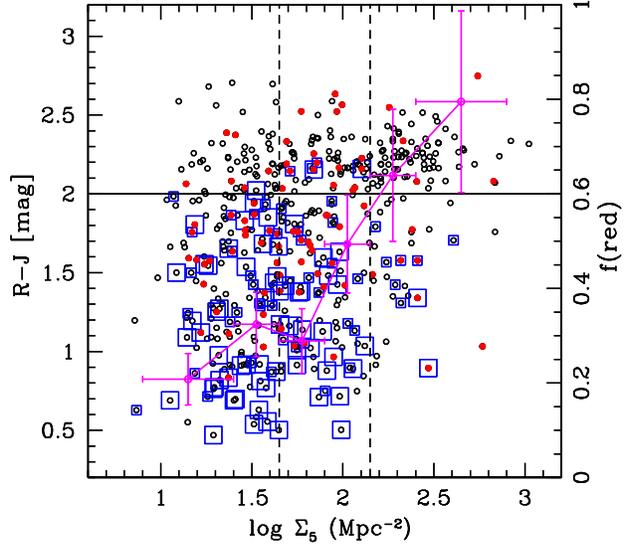}}
   \end{center}
   \vspace{-0.3cm}
\caption{The colour--density plot for all the member galaxies (open circles). 
  The small and large open squares indicate the weak (EW$<$50\AA) and strong 
 (EW$>$50\AA) H$\alpha$\ emitters, respectively.  The filled circles show 
  the MIR galaxies. The dashed lines at $\log \Sigma _{5} = 1.65$ and 
  $2.15$ indicate the dividing lines between the low-, medium- and 
  high-density regions as defined in Koyama et al.\ (2008). The 
  solid-line shows the locus of the fraction of red galaxies 
  calculated using our NIR-detected members.
  The vertical and horizontal error-bars indicate Poisson errors and the 
  size of the environmental bins used for f(red), respectively. }
\label{fig:color_density}
\end{figure}

As we showed in the previous sections, the nature of star-forming 
activity of galaxies is related to environment (i.e. galaxy density).
In Fig.~\ref{fig:color_density}, we plot $R-J$ colours as a function 
of the local density.  This plot would be a good summary of our
important findings. 

Firstly, colours of galaxies start to change at the medium-density 
environment. Clearly, the number of blue galaxies (with $R-J<2.0$)
decreases in high-density environment ($\log \Sigma _{5} > 2.15$).
This is quantified by calculating the fraction of red galaxies 
(with $R-J>2.0$) as a function of the local density (the solid-line 
locus in Fig.~\ref{fig:color_density}). 
This result is consistent with our previous finding in \cite{koy08},
who made a similar plot and examined $R-z'$ colour using all optical
sources. We confirm the trend that galaxy properties are changed not 
only in cluster environment but also in relatively low-density groups 
and filaments.

Secondly, we notice that most of the H$\alpha$\ emitters are blue, 
but there exist some red emitters with $R-J>2.0$. These galaxies are
candidates of dusty star-forming galaxies, and interestingly, these
galaxies are preferentially found in the medium-density environment.
In fact, we have 6 such red H$\alpha$\ emitters in total and 5 out of 6 are
in the medium-density environment. Also, 4 out of 6 such emitters 
are detected in MIR. This indicates that these red H$\alpha$\ emitters
are not just ceasing their star-formation. Rather, it seems that 
they have very strong star-formation but are heavily attenuated by dust.   

Thirdly, the MIR galaxies are generally redder than the H$\alpha$\ emitters,
and some of them are located on the red sequence. 
Such optically red MIR galaxies are also preferentially found 
in the medium-density region (i.e. transition environment) and some 
of them are not detected in H$\alpha$. 

It turns out that a fraction of the red galaxies
in the cluster surrounding regions are dusty star-forming galaxies,
which would have been difficult to be distinguished from
passively evolving red galaxies with optical colour information alone.
In contrast, most of the red galaxies in the high-density environment 
are passive galaxies with little on-going star-forming activity.
This may indicate that we are witnessing the transition phase of
galaxies in the cluster outskirts when/where some galaxies experience
an active, but dusty star-forming phase before being truncated.
More detailed discussion will follow in the next section.

\section{Hidden activity in distant clusters}

\subsection{Comparison between H$\alpha$\ and MIR activity}
As suggested in the previous section, a significant fraction
of star formation activity is likely to be hidden in the optical
data even in the H$\alpha$\ data, in particular in the surrounding
regions of high-$z$ clusters.
To quantify the amount of hidden star formation, we calculate the 
ratio of SFR(IR) to SFR(H$\alpha$) for our MIR-detected galaxies
down to 3.5$\sigma$ in detection ($\sim$ 10$M_{\odot}$/yr).
Here, we measure SFR(H$\alpha$) for the members that satisfy 
$J$$-$NB119$>$0.15 instead of $J$$-$NB119$>$0.3 that we used 
in Section~3.1, so that we can reach to fainter MIR-detected galaxies.
In Fig.~\ref{fig:sfr_comparison}, we plot the ratios, SFR(IR)/SFR(H$\alpha$), 
as a function of SFR(IR).  Here, we do not apply for any extinction correction
in deriving SFR(H$\alpha$).  We see a weak trend that galaxies with 
higher SFR(IR) tend to have higher SFR(IR)/SFR(H$\alpha$) ratios. 
This is consistent with previous studies which show that the dusty component
of star formation rate becomes higher as total star formation rate goes up
(e.g. \citealt{gea06}; \citealt{dop02}; \citealt{gar09}). 
We can also notice that red MIR galaxies (shown as filled symbols
in Fig.~\ref{fig:sfr_comparison}) tend to have higher 
SFR(IR)/SFR(H$\alpha$) ratios. This is reasonable because
their red colours can be the results of heavy dust attenuation.

The absolute values of SFR(IR)/SFR(H$\alpha$) span widely
from $\sim$2 to $\sim$40, depending on their SFR(IR).  
For moderately star-forming galaxies
(i.e. SFR(IR)$\sim$10$M_{\odot}$/yr), the extinction of H$\alpha$\ 
strength seems to be $\sim$ 1 mag, in excellent agreement with 
the study of local star-forming galaxies (e.g. \citealt{ken94}).
However, for actively star-forming galaxies (i.e. SFR(IR)$\gsim$
20$M_{\odot}$/yr), the magnitude of obscuration is significantly
large and $A_{\rm{H\alpha}}$ can exceed $\sim$3 mag in the extreme cases,
even though H$\alpha$\ line is considered as one of the most reliable SFR 
indicators.
There have been some pieces of evidence for such large amount
of obscuration in actively star-forming IR galaxies. 
For example, \cite{pog00} presented spectral natures of IR galaxies
at intermediate redshifts, and showed that 
these galaxies tend to be classified as 'e(a)', which show
emission lines as a sign of moderate star formation activity
and strong balmer absorption at the same time.
In fact, they showed SFR(IR)/SFR(H$\alpha$)$\sim$10--70,
and this is in good agreement with our values.
In cluster environments, \cite{sma99}
conducted deep radio observations in the CL0939 cluster at 
$z=0.41$. They detected radio continuum light at 1.4GHz,
as a sign of on-going star formation activity, from the
galaxies which have been classified as post-starburst galaxies
by optical spectra based on the presence of strong balmer absorption
lines and the absence of emission lines.
Recently, \cite{dre09} conducted MIR observation of the A851 cluster
at $z=0.41$.  They detected MIR emission not only from the majority of e(a)
galaxies but also interestingly from $\sim$30\% of k(a) galaxies
(i.e. post-starburst galaxies with no emission lines).
These results all indicate that the optical signs of star-forming
activity tend to be weak especially for galaxies with strong star formation
that are well traced by MIR observations, and thus can be sometimes misleading.

In fact, our MIR galaxies in RXJ1716 do not show strong H$\alpha$\ emissions.
As shown in Fig.~\ref{fig:SFR_vs_env}, we find that EW(H$\alpha$+[NII]) 
and SSFR(H$\alpha$) of the
MIR-detected galaxies tend to be lower than those of the MIR-undetected
H$\alpha$\ emitters.  Also, the MIR galaxies are sometimes 'not' detected
as H$\alpha$\ emitters. 
Unfortunately, we do not have any spectroscopic information for 
these galaxies yet.
It is possible that these galaxies are even more heavily obscured 
objects, but in the current situation, we cannot exclude the possibility that 
these galaxies are located at slightly lower or higher redshifts and their 
H$\alpha$\ lines do not fall within the NB119 filter.
Therefore, we limit the sample to the galaxies with H$\alpha$\ detection
in this paper because these galaxies are reliable members.  
We will present spectroscopic data of this cluster to reveal 
the nature of MIR galaxies undetected in H$\alpha$\ in a forth-coming paper.

\subsection{Environmental dependence of the hidden activity ?}

\begin{figure}
   \begin{center}
    \leavevmode
    \vspace{-1.7cm}
    \rotatebox{0}{\includegraphics[width=8.6cm,height=8.6cm]{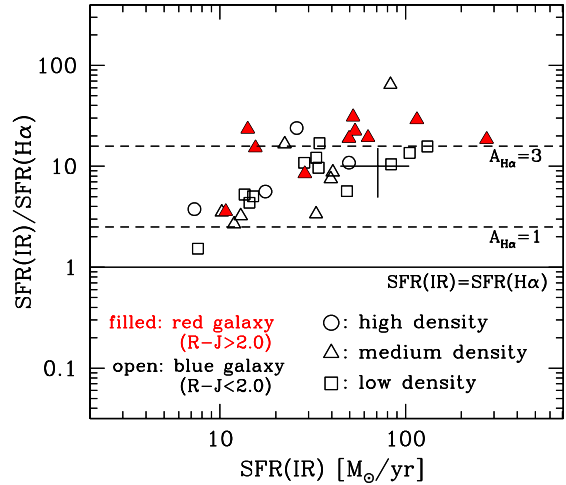}}
   \end{center} 
   \vspace{-0.5cm}
   \caption{The ratios of SFR(IR) to SFR(H$\alpha$) plotted against
  SFR(IR).  We plot only the H$\alpha$\ emitters with $J_{\textrm{corr}}$
  $-$NB119$>$0.15 that are detected as single objects at MIR with 
  detection significance of $>$3.5$\sigma$.
  The squares, triangles, and the circles indicate the galaxies
  in the low-, medium-, and high-density environments, respectively. 
  The filled symbols show the red galaxies with $R-J > 2.0$. 
  A representative error bar is shown as the cross. The uncertainty
  in SFR(H$\alpha$) corresponds to a 1$\sigma$ error in the aperture 
  photometry,
  while we adopt a 50\% error in SFR(IR) to take into account the large
  uncertainty in the conversion from $\nu L_{\nu , 8\mu \textrm{m}}$ to
  $L$(IR) (see text). }
\label{fig:sfr_comparison}
 \end{figure}

Taking the unique advantage of the wide-field coverages of our
both H$\alpha$\ and MIR imaging surveys presented in this paper,
we attempt to examine the environmental 
dependence of 'hidden' star-formation around a $z\sim 0.8$ 
cluster for the first time.
In Fig.~\ref{fig:sfr_comparison}, we use different symbols 
depending on their environment (squares for low-density, triangles
for medium-density and circles for high-density). 
Interestingly most of the 'heavily attenuated'
galaxies (e.g. SFR(IR)/SFR(H$\alpha$)$\gsim$15) are located
in the medium-density regions.
Although the number of MIR galaxies is not large, there is a hint
that very dusty galaxies most 
frequently appear in groups/filaments where galaxy properties
are dramatically changed (see Section~6.3). 
Note that some of those galaxies in the outskirts
may have large relative velocities with respect to the cluster core
and their H$\alpha$\ lines may fall on the wings of the transmission
curve of the NB119 filter, and therefore their SFR(H$\alpha$)
may be underestimated.  Their red colours, however, indicate that they
are indeed dusty objects. 
Also, it can be seen that bright MIR galaxies (e.g. SFR(IR)
$\gsim 40$$M_{\odot}$/yr) are found only in the medium- and low-density
environments, but no such galaxies are found in the high-density region.
These results are consistent with our findings in Section~6
where we find many dusty red galaxies in the medium-density regions.
Such enhanced dusty star-formation in the intermediate-density
environment has also been reported for lower-$z$ samples in the
literature (e.g. \citealt{wol05}; \citealt{fad08}; \citealt{gal09}; 
\citealt{tra09}).
Our current results show a similar trend up to $z\sim 0.8$.
Our unique wide-field H$\alpha$\ and MIR surveys in and around
the RXJ1716 cluster demonstrate the necessity of wide-field IR
observations covering the transition environment in the cluster
suburbs in order to reveal the key processes to shape the environmental
dependence of galaxy evolution and its link to IR activity of galaxies.

\section{Global evolution of star-formation activity in clusters}

\subsection{H$\alpha$\ view of the cluster evolution}
\begin{figure}
   \begin{center}
    \leavevmode
    \rotatebox{0}{\includegraphics[width=8.5cm,height=8.5cm]{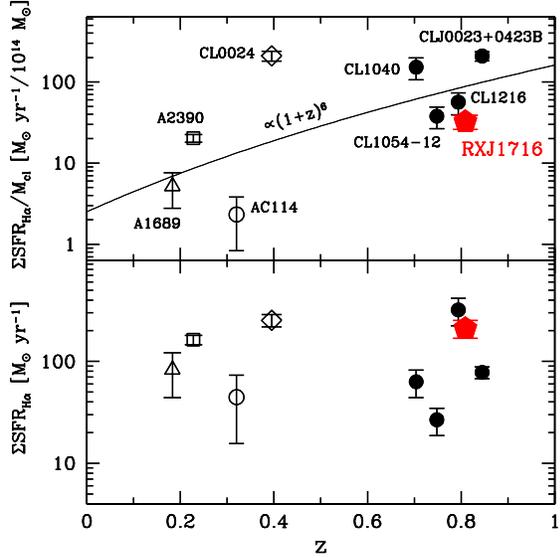}}
   \end{center} 
   \vspace{-0.5cm}
   \caption{ ({\it bottom}): The integrated total SFR(H$\alpha$) within
  0.5$\times$$R_{\textrm{200}}$ as a function of redshift. 
 ({\it top}): The integrated total SFR(H$\alpha$) normalized by the cluster
  mass $M_{200}$ as a function of redshift.  
 }
\label{fig:cluster_SFR_vs_z}
 \end{figure}

In this final section, we discuss the global evolution of star formation
activity in clusters together with the previous H$\alpha$\ cluster
studies in the literature.
\cite{fin05} sketched the evolution of H$\alpha$-derived total SF
in clusters ($\Sigma$SFR$_{\textrm{H}\alpha}$) and that divided by 
cluster mass ($\Sigma$SFR$_{\textrm{H}\alpha}$/$M_{cl}$) as a 
function of cluster redshift (see also \citealt{kod04}). 
We now estimate these quantities ($\Sigma$SFR$_{\textrm{H}\alpha}$ and 
$\Sigma$SFR$_{\textrm{H}\alpha}$/$M_{cl}$) for the RXJ1716 cluster.
\cite{fin05} used all the cluster member galaxies within 
0.5$\times$$R_{200}$ for each cluster to derive 
$\Sigma$SFR$_{\textrm{H}\alpha}$ and divide 
$\Sigma$SFR$_{\textrm{H}\alpha}$ by $M_{200}$ for each cluster
to derive $\Sigma$SFR$_{\textrm{H}\alpha}$/$M_{cl}$. 
The radius and mass ($R_{200}$ and $M_{200}$) of each cluster 
are defined by the following equations:
\begin{equation}
R_{200} = 2.47 \frac{\sigma}{1000 \rm{km/s}} \frac{1}{\sqrt{\Omega
 _{\Lambda} + \Omega _0 (1+z)^3}} \rm{Mpc}
\end{equation}
\begin{equation}
M_{cl} = 1.71 \times 10^{15} \left( \frac{\sigma}{1000 \rm{km/s}}  
\right)^3 \frac{1}{\sqrt{\Omega _{\Lambda} + \Omega _0 (1+z)^3}}
{M}_{\odot}
\end{equation}
where $\sigma$ is the velocity dispersion of the cluster.
However, for an unrelaxed cluster such as this RXJ1716 cluster,
the velocity dispersion tends to be over-estimated due to the
presence of kinematical substructures.
In fact, the velocity dispersion of the RXJ1716 cluster
is estimated to be $\sim$1500km/s in \cite{gio99},
which is much larger than that estimated from its X-ray
temperature ($\sim$600km/s) as they noted. 
Based on the new X-ray measurement in \cite{ett04}, we re-estimate
the velocity dispersion to be $\sigma =$837~km/s.
Based on this value, we derive $R_{200}$=1.3 Mpc and
$M_{200}$=6.5$\times$10$^{14} M_{\odot}$ for the RXJ1716 cluster.  
We then integrate SFR(H$\alpha$) of the member galaxies detected 
at more than 3$\sigma$ level in H$\alpha$\ within 0.5$\times$$R_{200}$, 
and correct for incompleteness due to the wings of the filter response
function by multiplying a factor $\sim$1.4 to derive 
$\Sigma$SFR$_{\textrm{H}\alpha}$.
We show our measurements, $\Sigma$SFR$_{\textrm{H}\alpha}$ 
$= 210 \pm 42 $ $M_{\odot}$/yr, in Fig.~\ref{fig:cluster_SFR_vs_z} with
those of other clusters in \cite{fin05} (i.e. originally from
\citealt{bal02} for A1689, \citealt{bal00} for A2390, \citealt{cou01} 
for AC114,
\citealt{kod04} for CL0024, \citealt{fin05} for CL1040, CL1054-12 and
CL1216, and \citealt{fin04} for CLJ0023+0423B).
As can be seen in Fig.~\ref{fig:cluster_SFR_vs_z}, we find no 
significant evolution in $\Sigma$SFR$_{\textrm{H}\alpha}$ with 
redshift (lower panel), 
while $\Sigma$SFR$_{\textrm{H}\alpha}$/$M_{cl}$ seems to 
increase towards higher redshift clusters (upper panel). 
Although the scatter at a given redshift is very large,
$\Sigma$SFR$_{\textrm{H}\alpha}$/$M_{cl}$ approximately scales 
with $(1+z)^6$ as shown by a solid curve in the top panel of 
Fig.~\ref{fig:cluster_SFR_vs_z}.

\begin{figure}
   \begin{center}
    \leavevmode
    \rotatebox{0}{\includegraphics[width=8.5cm,height=8.5cm]{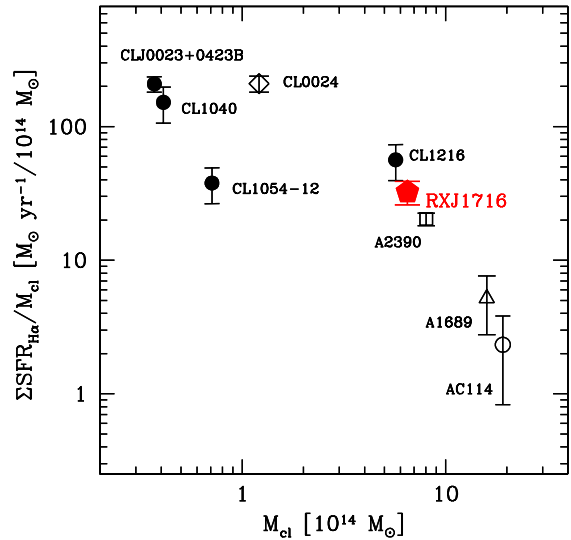}}
   \end{center} 
   \vspace{-0.5cm}
   \caption{The integrated total SFR(H$\alpha$) normalized by the cluster mass
  as a function of cluster mass, $M_{200}$. The symbols are the same as
  in Fig.~\ref{fig:cluster_SFR_vs_z}. Open and filled symbols indicate
  the low-$z$ samples ($z\lsim0.4$) and the high-$z$ samples ($z\gsim0.6$), 
  respectively.}
\label{fig:cluster_SFR_vs_Mcl}
 \end{figure}

It is possible that the scatter among clusters at a similar redshift
is due to the difference in cluster mass.
As shown in Fig.~\ref{fig:cluster_SFR_vs_Mcl}, 
$\Sigma$SFR$_{\textrm{H}\alpha}$/$M_{cl}$
is well correlated with cluster mass, in consistent with \cite{fin05}
(see also \citealt{pog06} and \citealt{hom05}). 
It may indicate that the cluster mass is also an important factor 
to determine cluster properties.
We should note, however, that the determination of cluster
mass is highly uncertain and this could be the biggest source of
uncertainty in drawing any general picture of cluster evolution. 
In fact, if we take $\sigma \sim 1500$km/s as derived from the optical
spectroscopy in \cite{gio99}, $\Sigma$SFR$_{\textrm{H}\alpha}$ 
increases by a factor of $\sim$2 and 
$\Sigma$SFR$_{\textrm{H}\alpha}$/$M_{cl}$ decreases by a 
factor of $\sim 3$. We would need a much larger sample of 
clusters to obtain more secure trend.


\subsection{MIR view of the cluster evolution}

In the previous sub-section, we have examined the evolution of 
star-forming activity in clusters based on H$\alpha$\ intensity
assuming a constant 1-magnitude extinction correction for all the emitters.
This assumption is valid for moderately or weakly star-forming galaxies, 
but much larger extinction correction would be needed for very actively 
star-forming galaxies (see Fig.~\ref{fig:sfr_comparison}).
In order to take into account the hidden star-formation as much as possible,
we now use SFR(IR) for the MIR-detected galaxies while we stick to use
SFR(H$\alpha$) with 1-mag correction for the MIR-undetected galaxies,
and sum up all the SFR within 0.5 $\times$$R_{\textrm{200}}$ from 
the cluster centre. In this calculation, we included the blended 
MIR sources as well, and used the combined MIR flux of such sources.
These all sum up to the total SFR(H$\alpha$+MIR) of $\sim 620\pm 250$ 
$M_{\odot}$/yr , which is more than factor 2 larger than the 
integrated SFR(H$\alpha$) even after applying for the constant 
1-mag extinction correction.
Note that we adopt a 50\% uncertainty in $\Sigma$SFR$_{\textrm{IR}}$,
which is the same amount as what we assumed in SFR(IR) for individual
galaxies (Section 2.3). This may be an overestimation for an integrated
quantity $\Sigma$SFR$_{\textrm{IR}}$, but we take it as a conservative error.

\begin{figure}
 \vspace{-1.5cm}
   \begin{center}
    \leavevmode
    \rotatebox{0}{\includegraphics[width=8.5cm,height=8.5cm]{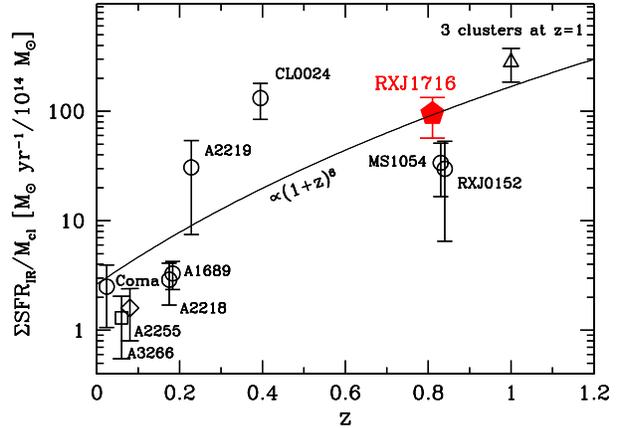}}
   \end{center} 
   \vspace{-0.5cm}
   \caption{The integrated total cluster SFR(IR) per unit cluster mass
  as a function of redshift.
  The open circles are taken from Bai et al.\ (2007). The open 
  square and the diamond are from Bai et al.\ (2009) and Shim et al.\
  (2009), respectively.  The open triangle shows the composite value 
  of 3 clusters at $z=1$ in Krick et al.\ (2009).   
  For RXJ1716, we combine the H$\alpha$\ data and the MIR data to derive 
  the total SFR (see text). }
\label{fig:cluster_SFRIR_vs_z}
 \end{figure}

We suggest that non-negligible fraction of star-forming activity in clusters
is hidden by dust and thus strongly supports the importance of IR study
of distant clusters and/or accurate extinction correction to obtain true
SFRs and to quantify the evolution of star-forming activity in clusters. 
Although the number of known clusters studied so far in MIR is very small,
some authors have attempted to sketch the global evolution of
star formation activity in clusters based on the MIR-derived SFRs
(\citealt{gea06}; \citealt{bai07}, 2009; \citealt{hai09a}; \citealt{kri09}). 
We here follow the scheme in \cite{bai07}, who measured integrated cluster
SFR(IR) within 0.5$\times$$R_{\textrm{200}}$ from the cluster centre.
Their flux limit in IR corresponds to $\sim$10$M_{\odot}$/yr at $z\sim0.8$,
but they integrated SFR down to $\sim$2$M_{\odot}$/yr by extrapolating
their IR luminosity function to that depth.
We can go down to the similar depth by combining the MIR- and 
H$\alpha$-derived SFRs within 0.5$\times$$R_{\textrm{200}}$ of the 
RXJ1716 cluster.  In Fig.~\ref{fig:cluster_SFRIR_vs_z}, 
we show the total SFR(H$\alpha$+MIR) normalized by cluster mass
for RXJ1716 cluster with SFRs(MIR) of other clusters in \cite{bai07}
(i.e. originally from MIR surveys by \citealt{bai06} for Coma,
\citealt{biv04} for A2218, 
\citealt{duc02} for A1689, \citealt{coi05a} for A2219, 
\citealt{coi05b} for Cl0024, \citealt{mar07} for RXJ0152 and
\citealt{bai07} for MS1054).  We also add three more recently 
reported data points at $z=0.06$ (A3266) from \cite{bai09}, at $z=0.08$ 
(A2255) from \cite{shi09} and at $z=1$ as a composite value of three 
$z\sim 1$ clusters presented in \cite{kri09}.  
We use $M_{\textrm{200}}$ as the cluster mass for RXJ1716 (same as in 
Section~8.1), but the trend does not change if we use a lensing mass
of $2.6\pm0.9 h^{-1} \times 10^{14}$ $M_{\odot}$ (\citealt{clo98}) 
to be consistent with \cite{bai07} who mainly use lensing masses.
We can see a similar evolutionary trend with redshift following 
$\propto (1+z)^6$in cluster star-forming activity to that seen 
in Fig.~\ref{fig:cluster_SFR_vs_z}. 

On the other hand, many field studies showed that the cosmic star
formation rate density or SSFR of field galaxies decline since $z\sim1$
following $\propto (1+z)^3$ (e.g. \citealt{hop04}; \citealt{yos06}). 
The difference in the power index between cluster and field suggests
that the galaxy evolution in cluster environment is accelerated compared
to that in the general field since $z\sim1$ (see also Kodama \& Bower 2001),
although we need a larger sample of distant clusters to confirm this
interesting result and to quantify environmental effects. 

It would be expected that the number (or fraction) of actively
star-forming galaxies and the relative importance of hidden
star-formation may increase with redshift.
The combined approach of deep H$\alpha$\ and MIR observations for
$z>>1$ clusters is thus essential and it will enable us to
quantify the importance of hidden side of star formation activity
in clusters and its evolution.

\section{Summary}
\label{sec:summary}

We conducted a narrow-band H$\alpha$\ imaging survey for the RXJ1716
cluster at $z=0.81$ with MOIRCS on the Subaru Telescope. This is the
first wide-field H$\alpha$\ emitter survey for $z\gsim 0.8$ clusters
to date, and we neatly covered the pre-defined filamentary large-scale
structures in and around this cluster.  Combining with the 
wide-field MIR imaging data taken with AKARI satellite, we mapped
out not only normal star-forming galaxies but also dusty starbursting
galaxies across a wide range in environment.
Our findings are summarized as follows:

\begin{enumerate}
\item The spatial distribution of the H$\alpha$\ emitters are very similar
to that of the MIR galaxies.  We find that the H$\alpha$\ emitters 
avoid cluster central region ($R_c \lsim$0.25Mpc) just as the MIR 
galaxies do, and that the fraction of H$\alpha$\ emitters is higher in the
lower-density environment. 

\vspace{2mm}

\item H$\alpha$\ emission lines are detected mainly from blue galaxies,
while the MIR galaxies tend to be redder. There are some MIR galaxies
on the red sequence, and interestingly, such red MIR galaxies 
are most commonly seen in the 'medium-density' 
environment such as cluster outskirts, groups and filaments.   
We also find that the MIR galaxies are {\it not} strong H$\alpha$\ emitters.
Also, there are some MIR galaxies without detectable H$\alpha$\ emission
lines (EW$\lsim$30\AA).
These suggest the existence of large amount of highly obscured
galaxies in the distant cluster environment.
However, we still need spectroscopic confirmation of membership of
the individual MIR and H$\alpha$\ sources to prove this interesting result.

\vspace{2mm}

\item We find some H$\alpha$\ emitters on the red sequence. Although
such population is rare (only 6 in total), these red H$\alpha$\ emitters 
are concentrated in the medium-density environment, similar to the spatial
distribution of MIR sources.
We also find that many of the red H$\alpha$\ emitters (4 out of 6) are 
detected in MIR.  This suggests that such red emitters are not just 
gradually truncating their star-forming activity but actively star-forming 
with heavy dust obscuration.

\vspace{2mm}

\item We examined the amount of hidden star formation based on the
ratio of SFR(IR)/SFR(H$\alpha$) for the MIR galaxies.
We find that for moderately star-forming galaxies with SFR(IR)$\lsim$
10$M_{\odot}$/yr the amount of extinction for H$\alpha$\ is $\sim$1 mag,
in excellent agreement with local spirals.  However, for actively 
star-forming galaxies with SFR(IR)$\gsim$20$M_{\odot}$/yr, H$\alpha$
lines are more heavily attenuated and the extinction can exceed
$A_{\textrm{H}\alpha}$$\sim$3 in the extreme cases.  Interestingly, most of
such very dusty galaxies are located in the medium-density environment,
suggesting the enhancement of star formation activity in such environment,
most likely triggered by galaxy-galaxy interactions and/or mergers.

\vspace{2mm}

\item Combining our unique H$\alpha$\ and MIR data, we derive total SFR
in the cluster down to $\sim$2$M_{\odot}$/yr within 
0.5$\times$$R_{\textrm{200}}$ from the cluster centre.
We find that the cluster total SFR based on H$\alpha$\ alone can be
underestimated about factor $\gsim$2 even after 1-mag extinction correction.
We suggest that the mass-normalized cluster SFR evolves with redshift
following $\propto (1+z)^6$, although the number of clusters studied so far
is still small.  We need a much larger sample of clusters studied in MIR
and should go back to more distant clusters at the same time, to fully
understand the evolution of star-forming activity in clusters.

\end{enumerate}

Overall, we have stressed the importance of cluster outskirts as a key 
environment for galaxy transition. 
Our results also imply that we need to be very careful when studying star 
formation activity in distant clusters.
We may miss a significant amount of star formation if we have only optical
information (even with the H$\alpha$\ line), in particular in the cluster
outskirts where starbursting populations are seen.
In order to trace the true star formation activity in distant clusters, 
MIR-FIR observations will be essential on top of the optical-NIR observations.

\section*{Acknowledgment}
We thank the anonymous referee for the careful reading of the
paper and for helpful suggestions, which improved the paper. 
We thank Dr. Hyunjin Shim and Dr. Myungshin Im for kindly 
providing us their data for A2255. 
The optical and NIR data used in this paper are collected at the 
Subaru Telescope, which is operated by the National Astronomical 
Observatory of Japan (NAOJ). 
We acknowledge Dr. Masami Ouchi and Dr. Ryuji Suzuki for their
great contribution to designing the NB119 filter on MOIRCS.
The mid-infrared data that we used in this study are based on 
observations with AKARI, a JAXA project with the participation with ESA.  
This work was financially supported in part by a Grant-in-Aid for the
Scientific Research (Nos.\, 18684004; 21340045) by the Japanese 
Ministry of Education, Culture, Sports and Science. 
Y.K. and M.H. acknowledge support from the Japan Society for the Promotion 
of Science (JSPS) through JSPS research fellowships for Young Scientists.


\end{document}